\documentclass[useAMS,usenatbib]{mn2e}

\usepackage{epsfig}
\usepackage{lscape}

\newcommand{\lsimeq}{{_<\atop^{\sim}}}

\newcommand{\aj}{AJ}
\newcommand{\nat}{Nature}
\newcommand{\apj}{ApJ}
\newcommand{\apjl}{ApJL}
\newcommand{\apjs}{ApJS}
\newcommand{\aap}{A\&A}
\newcommand{\aaps}{A\&AS}
\newcommand{\mnras}{MNRAS}
\newcommand{\aapr}{A\&ARv}
\newcommand{\pasj}{PASJ}
   
\title[Hidden Activity in High-redshift Spheroidal Galaxies]
{Hidden activity in high-redshift spheroidal galaxies from
  mid-infrared and X-ray observations in the GOODS-North field}

\author[G. Rodighiero et al.]
       {G. Rodighiero$^1$, C. Gruppioni$^2$, F. Civano$^{3,2}$, A. Comastri$^2$, 
       A. Franceschini$^1$,  \newauthor M. Mignoli$^2$, J. Fritz$^1$,
       C. Vignali$^3$, T. Treu$^4$ \\
        $^1$ Dipartimento di Astronomia, Universit\`a di Padova, vicolo dell'Osservatorio 2, 
             I--35122 Padova, Italy\\
        $^2$ INAF -- Osservatorio Astronomico di Bologna, via Ranzani 1, I--40127 Bologna, Italy\\
        $^3$ Dipartimento di Astronomia, Universit\`a di Bologna, via Ranzani 1, I--40127 Bologna, Italy\\
        $^4$ Physics Department, University of California, Santa Barbara, CA 93106-9530}
\begin{document}

\maketitle

\begin{abstract}
We exploit very deep mid-IR (MIR) and X-ray observations by $Spitzer$ and $Chandra$ in the
GOODS North field to identify signs of hidden (either
starburst or AGN) activity in spheroidal galaxies between $z\simeq 0.3$ and 1.
Our reference is a complete sample of 168 morphologically classified spheroidal 
(elliptical/lenticular) galaxies with $z_{AB}<22.5$ selected from GOODS ACS imaging.
Nineteen of these have 24-$\mu$m detections in the GOODS catalogue, half of which 
have an X-ray counterpart in the 2 Ms $Chandra$ catalogue (6 detected in the 2-10 
keV X-rays hard band), while about 25\% have 1.4 GHz fluxes larger than
40 $\mu$Jy. 
Traces of hidden activity in the spheroidal population are also searched for
in the deep X-ray images and 14 additional galaxies are detected in X-rays only.
The nature of the observed MIR emissions is investigated by modelling their Spectral 
Energy Distributions (SEDs) based on the available multi-wavelength photometry, 
including X-ray, UV, optical, near-IR, MIR and radio fluxes, and optical spectroscopy. 
The amount of dust derived from the IR emission observed by $Spitzer$ 
appears in excess of that expected by mass loss from evolved stars. 

When the available independent diagnostics are compared, in general they provide
consistent classifications about the nature of the activity in our
spheroidal population. Given that, in principle, none of these
diagnostics alone can be considered as conclusive, only trough a panchromatic
comparison of them we can reach an accurate comprehension of the
underlying physical processes.     
In particular, our multi-wavelength analysis of the X-ray and MIR properties leads us to 
conclude that at least 8 of the 19  24-$\mu$m bright sources should hide an obscured AGN, while
the X-ray undetected sources are more likely dominated by star formation.
We conclude that $\sim$30 objects ($\sim$20\%) of the original flux-limited sample of 168 spheroidal 
galaxies in the GOODS-North are detected during phases of prominent activity, of both stellar and quasar 
origin.
Due to the short expected lifetimes of 
the IR and X-ray emissions, this fraction might imply a significant level of 
activity in this class of galaxies during the relatively recent cosmic
epochs -- $z\sim 0.3$ to $z\sim 1$ -- under investigation.

\end{abstract}


\section{Introduction}
\label{intro}

The origin of spheroidal (elliptical and lenticular) galaxies has
been, and still partly remains, a controversial issue.
Published results from high redshift galaxy
surveys appear not unfrequently in disagreement with each other. 
One example is the apparent conflict between
reports of the detection of massive elliptical galaxies at very high
redshifts (e.g. Glazebrook et al. 2004; Cimatti et al. 2004; McCarthy
et al. 2004; Daddi et al. 2005; Saracco et al. 2005) and the 
indications for a decline in their comoving number densities at
$z>1$ (e.g. Franceschini et al. 1998; Fontana et al. 2004; Faber et al. 2005). 

Some recent progress in this field has been achieved in particular by the
combined use of the IR multi-wavelength coverage offered by the Spitzer Space
Telescope, the unique imaging capabilities of HST/ACS, and the
enormous photon-collecting power of spectrographs on large
ground-based telescopes. This wealth of data significantly refined our 
knowledge of the build up of galaxy populations and revealed 
evidence of "downsizing" in galaxy formation, i.e. star formation
ending first in massive galaxies than in lower mass ones, as observed 
both at low and high redshift (Cowie et al.
1996; Tanaka et al. 2005; van der Wel et al. 2005; Treu et al. 2005; 
Juneau et al. 2005; Bundy et al.
2006; Borch et al. 2006; Jimenez et al. 2006). 

In particular, Treu et al. (2005) studied the evolution of the fundamental plane 
as a diagnostic for star formation and mass assembly history of
early-type galaxies. They suggested that most of the stellar mass in spheroidal
galaxies formed at $z>2$, with subsequent activity continuing to
lower redshifts ($z<1.2$). The fraction of stellar mass formed at
recent times depends strongly on galactic mass, ranging from $<1$\% for masses 
above $10^{11.5}\ M_{\odot}$ to 20\%-­40\% below $10^{11}\ M_{\odot}$.

Various approaches have been followed in order to detect recent
or ongoing activity in spheroidal galaxies to $z\sim1$. 
One approach was to look for close companions or morphological 
disturbances, and how they evolve as a function of redshift 
(Le Fevre et al. 2000; Patton et al. 2002; Khochfar \&
Burkert 2003; Lin et al. 2004; Cassata et al. 2005), under the
assumption that the physical mechanism triggering the activity and
responsible for the progressive build-up of galaxies are
interactions and merging. 
Another way of attacking the problem was to consider optical
colour information and look for blue ellipticals to find the
presence of young stellar populations, hence recent star-formation, e.g.
triggered by merging (Hogg et al. 2003; Bell et al. 2005; Cassata et al. 2006). 
As an example, Treu et al. (2005) provided independent support for recent activity in
spheroidals up to $z\sim1$ through spectroscopic ([O II] emission, H$\delta$) and
photometric (blue cores and broadband colors) diagnostics. 

An important related question is how the observed galaxy build-up
matches with evidences about the growth of nuclear super-massive
black-holes (BHs) during the quasar and AGN phase. A close
relationship of the two is suggested by the observed
correlations of the BH mass with the host galaxy mass and central
velocity dispersion in the local universe (Ferrarese \& Merritt 2000)
and by the concomitant presence of star-formation and
AGN activity sometimes directly observed in forming galaxies 
(Alexander et al. 2005). 
Many have suggested that AGN may be the cause or at least related to the
quenching of galaxies (see Silk \& Rees 1998, Croton et al. 2005, Hopkins
et al. 2006, among others). 


We follow in this paper an alternative line of investigation on high-$z$ ($z \sim 0.7$) 
spheroidal galaxies by looking for signs of optically hidden activity 
in their mid-infrared (MIR) and hard X-ray emissions (Rigby et al. 2004,
Georgantopoulos et al. 2006).  
Ongoing star-formation and young massive stars in local galaxies appear mostly embedded in dusty,
optically thick molecular clouds. 
Related to this, the {\em Infrared Space Observatory (ISO)} has shown
the presence of Polycyclic Aromatic Hydrocarbon molecules (PAHs) in
the spectra of some local early-type galaxies (i.e. Madden et
al. 1999; Xilouris et al. 2004). These results have been recently
confirmed by observations with the {\em Spitzer Space Telescope} (Pahre et
al. 2004, Bressan et al. 2006). However, the origin of the dust
responsible for the MIR emission in early-type galaxies could be
attributed either to merging events (Xilouris et al. 2004), cooling flows (Fabian et al. 1991) or to 
mass loss from late-type stars (Knapp et al. 1992).

In addition, a significant (even major) fraction of AGN accretion is
also expected to be optically obscured and to emerge in the MIR.
Then an effective way of identifying hidden star-formation and/or AGN 
activity in high-redshift spheroidal galaxies is to exploit deep mid- and far-IR photometry
with Spitzer MIPS on sky areas where morphological classification is
possible from deep HST/ACS high-resolution imaging. 

Hard X-ray observations, in turn, provide an even more straightforward
way to identify obscured AGNs, since hard X-rays are typically produced 
in copious amounts by them, and are able to penetrate the dust absorbing the
optical, UV and soft X-ray light emitted by the nucleus. In addition,
when the sensitivity of the X-ray data is high enough, they bear
potential independent information on deeply embedded young stellar
populations (Ranalli et al. 2003; Franceschini et al. 2003). 
Therefore, deep hard X-ray data combined with MIR
information can provide extraordinary tools for unveiling hidden AGNs
and starbursts in morphologically classified spheroidal galaxies and
to constrain the nature of these emissions. 

Many recent papers have studied the presence of AGN in high-$z$ early-types
(e.g. Georgakakis et al. 2006, Grogin et al. 2005, and Brand et
al. 2005).  Similar works at lower redshift are based on the Sloan Digital
Sky Survey (SDSS, Stoughton et al. 2002) that study AGNs in high-mass
early-type galaxies (Miller et al. 2003, Kauffmann et al. 2004).
More recently, Capetti and Balmaverde (2005, 2006 among the others) explored the
connection between the multiwavelength properties of AGNs in nearby
early-type galaxies and the characteristics of their hosts. 
Also Treu et al. (2005) found that $\sim$5\% of their E+S0 galaxies
are identified as AGNs, based on an X-ray luminosity above $10^{42}$ erg/s. 
An additional $\sim$10\% of the spheroidal sample is also detected with X-ray luminosities
between $10^{40}$ and $10^{42}$ erg/s, which are interpreted as mostly
due to a low-luminosity AGN or to ongoing star formation.

In this context, the GOODS-North (GOODS-N), with its deepest
available MIR (from $Spitzer$) and deeper hard X-ray (from $Chandra$)
data, offers a unique opportunity of investigating activity phenomena
in spheroidal galaxies. 

We focus in this paper on the specific comparison between the Chandra X-ray
view and the Spitzer 24 $\mu$m  view of AGN and star-formation
activity (Rowan-Robinson et al. 2005, Franceschini et al. 2005),
investigating in particular if the Chandra and Spitzer indicators agree and
correlate with one another when they are combined. 

Based on a flux-limited optical galaxy sample ($z_{AB}<22.5$) 
by Bundy et al.(2005), including 168 morphologically classified
elliptical/lenticular (E/S0) galaxies in the GOODS-N, we have identified
those with a MIR counterpart in the MIPS 24 $\mu$m GOODS catalogue
($S_{24\mu m}> 80\mu Jy$). 
We have also cross-correlated the optically-selected sample with 
the 2 Msec X--ray image of {\it $Chandra$ } reported by Alexander 
et al. (2003). For these galaxies we have built a multi-wavelength 
photometric catalogue including IRAC data at 3.6, 4.5, 5.8 and 8 $\mu$m 
(from the public GOODS release), 
Subaru and HST $U,B,V,R,i,z$ and $HK^\prime$ data
(Capak et al. 2004), and radio VLA 1.4GHz data (Richards et al. 2000).
Spectroscopic redshifts are available for all the sources of our sample. The galaxies are found at 0.1
$\lsimeq z \lsimeq$ 1.2, with a median redshift $<z>=0.68$.   
About half of the spheroidal galaxies with MIR emission present an 
X-ray counterpart and one-fourth have radio detection. 

We note here that half of our sample overlaps the spectroscopic sample discussed by
Treu et al. (2005). 
  
The present paper is structured as follows. Section 2 describes the multi-wavelength 
data-set and Section 3 our galaxy sample selection. In Section 4 we give 
details about our data analysis, by exploiting in particular the SEDs of our 
galaxies, the information from multi-colours and broad-band flux ratios.
Section 5 discusses and critically analyses these results, while Section 
6 summarizes our conclusions. 

We adopt a cosmological model with $\Omega_m$=0.3,
$\Omega_{\Lambda}$=0.7,  and $H_0=70~ km~s^{-1}~Mpc^{-1}$.

\section{THE DATA}
\label{sample}

With the aim of studying the multiwavelength photometry of extragalactic sources
measured by different instruments, we need to measure the bulk
of the emission from each object in each photometric band.
Only with such kind of approach an SED can be considered reliable. 
In our work, we performed aperture photometry in most bands, and decided to bring the
measurements as close to the total magnitudes as possible (see also
Yan et al. 2005).  

\begin{figure*}
\centerline{\psfig{figure=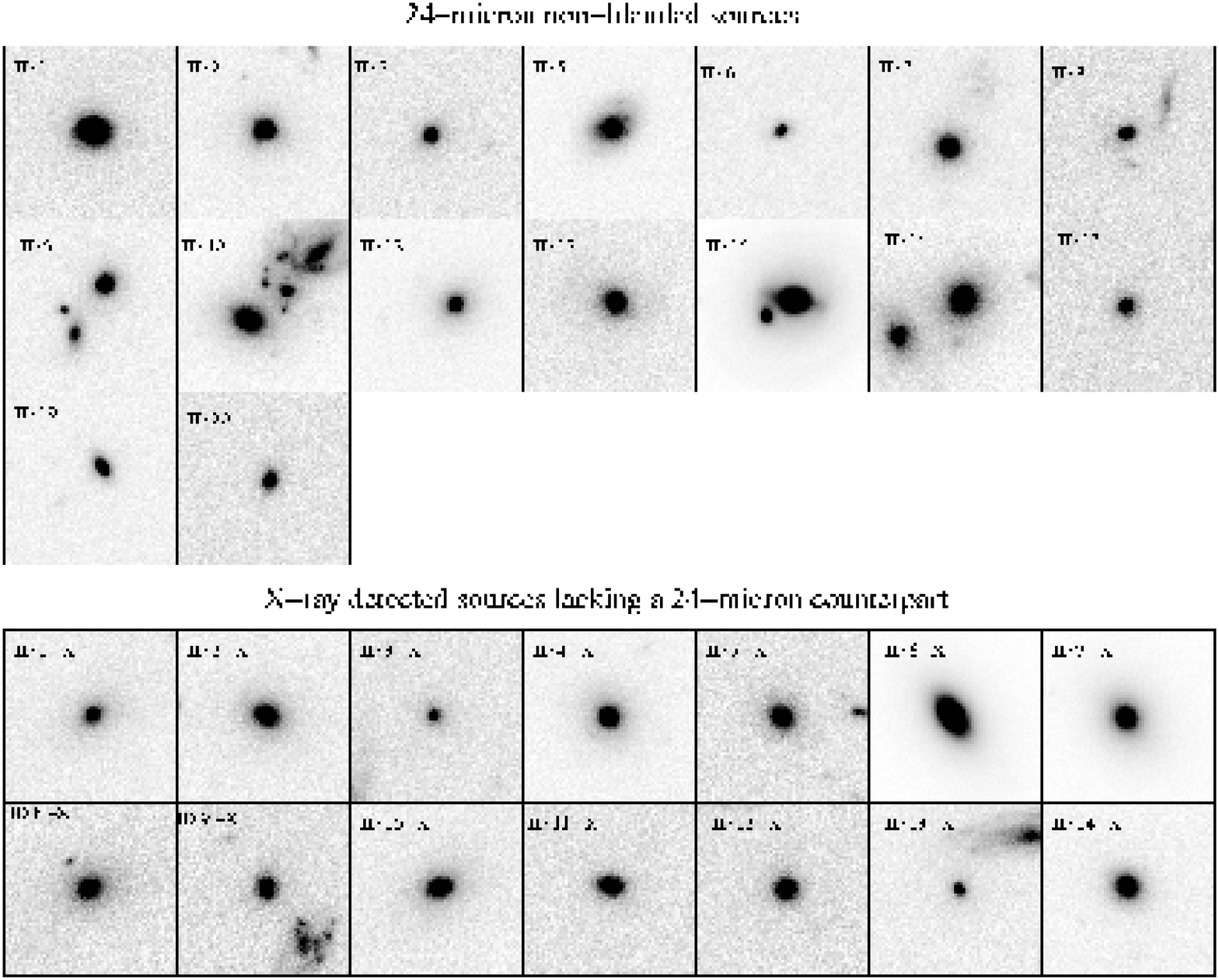,width=20cm}}
\caption{Top panel: optical images of the 16 spheroids with MIR emission and reliable
optical associations. Bottom panel:  optical images of the 14
spheroids with X-ray emission and undetected in the MIR. 
For each source a cutout of the z' band
ACS images is presented. The identification numbers are
those reported in Table 3. Each postage has a size of
8''x8''. North is up, East is at left.} 
\label{s_morph}
\end{figure*}

\subsection{Optical Imaging}
\label{optical}
The GOODS project (Dickinson et al. 2003)
has surveyed with ACS/HST two separate fields, the $Chandra$  Deep Field South 
(CDFS) and the Hubble Deep Field North (HDFN) with
four broad-band filters: F435W (B), F606W(V), F775W(i), and F850LP(z). 
The analysis presented here is based on the version 1.0 of the reduced, stacked 
and mosaiced images for all the data acquired over the five epochs of 
observation (Giavalisco et al. 2004). 

Source extraction and photometric measurements in the B, V, $i$ and $z$ bands 
have been performed by running a modified version of SExtractor (Bertin
\& Arnouts, 1996). 
We have considered the total magnitudes (as from the BEST$\_$MAG output 
parameter in SExtractor, see Giavalisco et al. 2004). 

Moreover, optical and near-IR ground-based photometric imaging in several bands 
for the GOODS-North field was taken from Capak et al. (2004).
The $U$-band data were collected using the Kitt Peak National
Observatory (KPNO) 4m telescope with the MOSAIC prime focus camera.
The $B$-, $V$-, $R$-, $I$-, and $z'$-band data were collected using the Subaru
8.2 m  telescope and Suprime-Cam instrument (Miyazaki et al. 2002).
The HK' data were collected using the QUIRC camera on the University
of Hawaii  2.2 m telescope (Hodapp et al. 1996). Details on the data 
reduction and photometric analysis are reported in Capak et al. (2004).   
A 3'' diameter aperture was used for the photometry. We applied the
corresponding aperture corrections reported by Capak et al. in all bands. 

\subsection{Morphological Analysis}
\label{morph}

Our basic selection is a sample of morphologically classified
elliptical galaxies in the GOODS-N based on the imaging data. 
We used the publicly available catalogue of Bundy et
al. (2005){\footnote {available at http://www.astro.caltech.edu/GOODS\_morphs}},
which was constructed with a magnitude limit of $z_{AB}=22.5$
over an area of 160 $arcmin^2$. The morphological selection is based 
on the visual inspection of ACS version 1.0 data.
Our sample selection is based on a catalogue of 168 {\sl bona-fide} 
E/S0s (classes 0 and 1 in Bundy's catalogue) out of a total sample of
1576 galaxies brighter than this magnitude limit.

\subsection{Deep $Spitzer$ Imaging in the Near- and Mid-IR}
\label{mips}

As part of the GOODS project, the $Spitzer$ Space Telescope has surveyed the 
GOODS-N field in the IR between 3.6 and 8.0 $\mu$m using IRAC and in the range 
24-160 $\mu$m with MIPS. In this paper we have made use of the public reduced 
data released by the GOODS team and available on the 
Web{\footnote {http://data.spitzer.caltech.edu/popular/goods}}.

For the IRAC data, we have adopted the official GOODS reduced images
(epoch 1 + 2) and performed our own photometry.
The IRAC source identification has been performed with SExtractor
independently in the four maps at 3.6, 4.5, 5.8 and 8.0 $\mu$m.

In order to obtain the most accurate Spectral Energy Distributions (SEDs),
we have computed the fluxes of each source by performing aperture photometry
in the four IRAC bands at the positions originally detected in the ACS 
$z$-band. Assuming that essentially all the sample sources 
are seen as point-like by the IRAC $\sim 2\ arcsec$ 
FWHM PSF imager, we computed with SExtractor (Bertin \& Arnouts, 1996) 
the fluxes within a 3.8 arcsec diameter aperture. 
This choice is supported by an accurate analysis performed by the SWIRE
team{\footnote {http://data.spitzer.caltech.edu/popular/swire/20050603\_enhanced\_v1/}}. 
They constructed color-magnitude diagrams for various types of
objects, in particular main-sequence stars. It was then found that the
scatter in these diagrams is minimized through the use of an aperture
of 3.8 diameter aperture, and corresponds to roughly twice the
beam-width. To obtain total fluxes, we then applied the correction 
factors indicated by the SWIRE team{\footnote {See Note 2}}. 
We have independently verified that the IRAC/SWIRE aperture
corrections are consistent with those derived by fitting the radial
brightness profiles of few stars in the GOODS fields (bright stars where selected
on the basis of their visual morphology in the ACS bands). 
For the only one source spatially resolved (\# 15), we used Kron like 
magnitudes (AUTO$\_$MAG output parameter in SExtractor).
For three blended IRAC sources present in our sample, we applied a
deconvolution procedure, as described in Section 3.1.

The public MIPS data-set includes calibrated maps and a catalogue of 24 $\mu$m 
sources with flux densities $S_{24}>80 \mu$Jy. The public photometry is based 
on a PSF fitting algorithm, where the SExtractor positions of the IRAC 
sources are used as input to the MIPS source extraction process.
The MIPS 24 $\mu$m PSF was generated from isolated sources in the
image, and re-normalized based on the aperture corrections published in
the MIPS Data Handbook (v2.1, section 3.7.5, table 3.12).

IRS Peak-Up imaging data at 16 $\mu$m are also available from Teplitz
et al. (2005) over an area of 35 square arcminutes in the GOODS-N to an
average 3 sigma depth of 0.075 mJy, for a total 149 detected sources. 
Three of these are in common with our spheroidal galaxy sample.
The 16 $\mu$m photometric band provides unique information on the
shape of the observed SEDs. In fact, the combination of the complete Spitzer filters set
represents a powerful tool to sample the evolution of the PAH features. 
For example, the 7.7 $\mu$m PAH feature can be traced from the local
universe,  where it lies in the IRAC channel 4 passband, to $0.8<z<1.3$,
where it falls in the 16 $\mu$m IRS Peak-Up filter,
up to redshifts near $z<2$, where it enters the 24 $\mu$m MIPS filter
(Teplitz et al. 2005).

\begin{figure}
\psfig{file=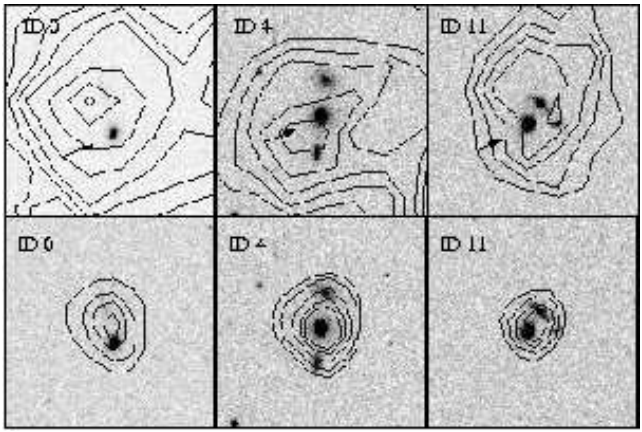,width=8.5cm}
\caption{Overlays of the MIPS 24 $\mu$m  contours over the optical ACS i' band images 
for spheroidal galaxies with multiple IDs inside the 24$\mu$m error-box 
(blended candidates). From upper left panel to right panel
ID \#0 (z=0.41), ID 4 (z=0.845), ID 11 (z=0.766). The lower panel corresponds to
overlays with the IRAC 3.6$\mu$m contours.
Each postage has a size of 8''x8''. North is up, East is to the left. }
\label{blend}
\end{figure}

\subsection{X-ray data}

Very deep X--ray observations (2 Megaseconds in total) of the {\it $Chandra$} Deep Field North 
(CDFN) have been reported by Alexander et al. (2003). We used them to
search for nuclear activity in spheroidal galaxies of the GOODS-North field 
in addition to those obtained from the IR emissions. 
The X--ray data have been retrieved from the public archive
and processed with standard tools making use of the calibrations
associated with the {\tt CIAO}\footnote{http://cxc.harvard.edu/ciao/}
software (version 3.2.1). Each of the 20 pointings of the $Chandra$  dataset
has been registered on a reference pointing (OBSID 3293) and aligned using 
the {\tt align\_evt}\footnote{http://cxc.harvard.edu/cal/ASPECT/align\_evt/} 
tool.  The charge transfer inefficiency and gain corrections were also 
applied to each single pointing. 
X--ray images were accumulated in the soft (0.5--2.0 keV), hard (2--8 keV) 
and full (0.5--8 keV) bands. 

X--ray counts in the three bands were extracted within a
circular region whose radius is a function of the off--axis angle
(typically of the order of 2--3 arcsec).  The background was estimated
locally for each source as the average of counts in several source--free 
surrounding regions.  The number of counts and the X-ray fluxes are fully consistent with the
values published in Alexander et al. (2003), and are reported in
Table 3.

\subsection{Optical spectroscopy}
\label{spec}

In the last few years various observational programs 
have undertaken a systematic spectroscopic follow-up in the GOODS-N.
In particular, the Keck team (Wirth et al. 2004) reported the results
of an extensive imaging and spectroscopic survey in the GOODS-North field
with DEIMOS on the Keck-II telescope. Observations of 2018
targets from a magnitude-limited sample of 2911 objects to $R_{AB}$ = 24.4
yielded secure redshifts for a sample of 1440 galaxies and AGNs, plus 96 
stars.  All the Keck spectra are publicly available{\footnote 
{http://www2.keck.hawaii.edu/science/tksurvey}}.
All our objects have a redshift from the Wirth et al. (2004)
catalogue. However, we did not found the corresponding spectrum in the
Keck archive for 10 sources in our sample (IDs 3, 4, 9, 10, 15, 16 and
IDs 3x, 4x, 8x and 10x as in Table 3). 
We then recovered four of the missing spectra (IDs 4, 9, 10 and 16)
from the sample of Treu et al. (2005).
All sources have a z-quality value of 4 (99\% confidence level) or 3
(90\% confidence level), as in the code assigned by Keck team. 

\begin{figure*}
\psfig{file=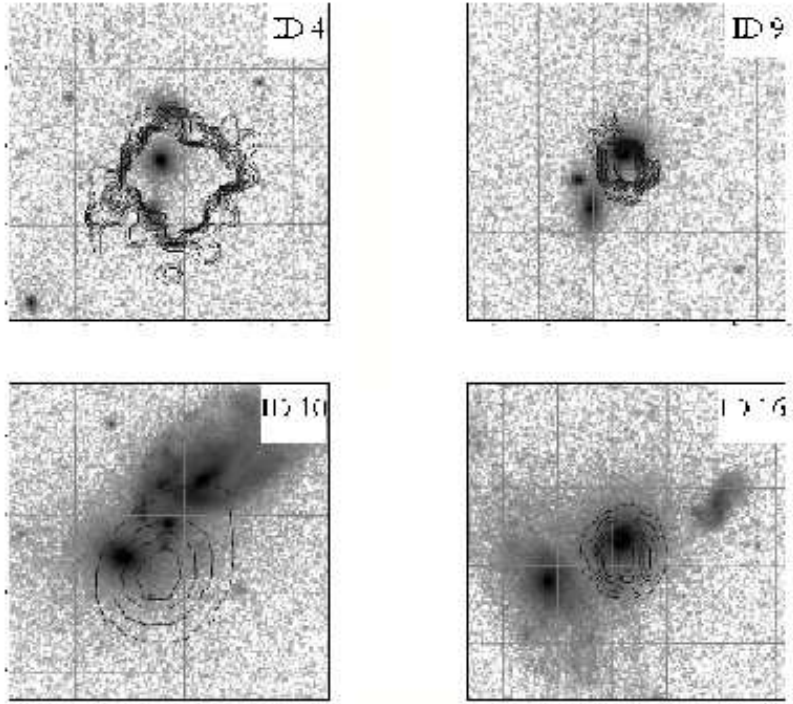,width=14cm,angle=0}
\caption{Overlays of the \textsl{i-}band images with X-ray \textsl{Chandra} 
contours for 4 spheroidal galaxies with confused identifications for the
MIPS 24 $\mu$m source.}
\label{x_acs}
\end{figure*}

\section{SAMPLE SELECTION}
\label{sample}

\subsection{Spheroids with mid-IR excess emission}
\label{spheroid}

As a first step, we cross-correlated the optical
catalogue including the 168 morphologically selected spheroids (see
Section \ref{morph}), with the bright $Spitzer-MIPS$ 24 $\mu$m catalogue
provided by the GOODS team ($S_{24}>80 \mu$Jy, Section \ref{mips}).
For 21 of the 168 spheroids we found a MIR counterpart within
a distance of 2'' from the optical position.
We checked  the morphology of each of the 21 
candidates taken from the Bundy et al. (2005) catalogue on the $i$-band ACS image.
We then excluded two objects (\#14 and \#18) from the sample  because they look
like late-type galaxies rather than E/S0 to our visual inspection
(however, we have verified that the inclusion of these two sources
does not change the main conclusions of this work).  


Source confusion is a potential problem for the optical 
identification of the 19 MIR sources in the final sample: the very different angular
resolution of the $Spitzer$ and HST instruments enhances the probability 
of finding more than one ACS source falling within a MIPS beam. 
However, for the majority of our sources the optical to NIR 
and MIR identifications turned out to be straightforward.
Only six of the optically-selected galaxies present extended 
IRAC or MIPS emissions and confusion problems. For three of
them (\#9, \#10 and \#16) we could use the spatial
information available in the HK'-band image to deconvolve the
3.6-8.0 $\mu$m fluxes (for details on this procedure see 
Franceschini et al. 2006).
Using  the HK' positions for reference, we applied a PSF-fitting algorithm
based on IDL procedure written ad-hoc to deblend also the 24 $\mu$m emission.

The confusion issue for the three remaining objects (ID \#0, \#4 and \#11) 
cannot be resolved by just looking at the HK' image. 
We will consider these 3 $blended$ sources separately, and their MIR
fluxes as upper limits of the spheroidal galaxy emission,
while the final catalogue of robust unambiguous associations includes 16 E/S0s
with 24 $\mu$m emission above $S_{24}>80 \mu$Jy.

Figure \ref{s_morph} (upper panel) shows optical postage stamps of the 16 reliable 
spheroids in our final sample. For each source a cutout of the four ACS images 
(B, V, i' and z' bands) is presented. The identification numbers are
those appearing in Table 3 (see also \S \ref{multicat}).
Postage stamps of the three blended sources are reported in Figure \ref{blend}, 
where the optical ACS i' band images are overlaid on the MIPS 24 $\mu$m
contours.   For the 16 isolated spheroids, the association of the
24 $\mu$m and the optical/near-IR source is unambiguous.

For what concerns the sources affected by confusion problems, the X-ray
counterparts, where available, (see Figure \ref{x_acs} below) 
support the hypothesis that the spheroidal object is strictly correlated 
with the stronger IR emitter, under the assumption that the two emissions
are related.


%
%

\subsection{Spheroids with X-ray emission}
\label{X}

More than half (12/19) of our {\it Spitzer}--MIPS sources 
have a unique X--ray counterpart within a search radius of 2 arcsec.
Four of the 12 X--ray sources are associated with the objects showing
confusion problems at 24 $\mu$m (see $\S$ \ref{spheroid}).
By exploiting the arcsec X--ray 
image quality, the soft X--ray contours  of the 4 {\it $Chandra$ } 
sources were overlaid on the HST/ACS images (Figure \ref{x_acs}). 
The X--ray/HST association is obvious for 3 of them 
where the X--ray emission is centered on the spheroidal component. 
The closest counterpart of source 10 is also a spheroidal galaxy,
although the correct association is not so clear.

For six out of the seven MIR spheroidal galaxies undetected 
in both soft and hard X--ray bands, upper limits were computed by 
X--ray counts in circular regions (2 arcsec of radius) 
of the CDFN map centered and reported in Table 3.
The optical position of source 19 was too close to a bright X--ray source
to obtain a reliable estimate of the X--ray upper limit.

In addition, we found 14 associations of 
Chandra X-ray sources among the 147 $z$-band selected spheroidal 
galaxies in the GOODS-North area without detectable MIR emission.
The optical z-band images for all of them are reported in Figure 
\ref{s_morph} (lower panel). All appear to display standard elliptical morphologies.
Photometric and other data for this sub-sample are reported in the
third panel of Table 3. The majority
of these sources (8 out of 14) are detected in hard band (2-8 keV).

To summarize, these 14 additional X-ray luminous objects 
bring our total sample of spheroidal galaxies in the 160 sq. arcmin 
GOODS-North field with signs of activity (either from MIR
or X-ray data) to a total of 33 out of 168 objects in the original 
optical sample (corresponding to a fraction of 20\%).

\subsection{Summary of the multiwavelength photometry for spheroids in the GOODS-N}
\label{multicat}

We have built a multi-wavelength photometric catalogue for our sample of 
spheroidal galaxies with MIR or X-ray emissions in the following way. 
Starting from the optical $z$-band positions, we used a search radius of 1'' in
each bands to look for the cross-identifications in the corresponding catalogues.
As mentioned in Section \ref{spheroid}, direct visual inspection of the
ACS images guarantees the reliability of the final counterparts. 

The photometry for each source is presented in Table 3.
We adopt a common value of 10\% (15\%) of the measured fluxes as 
photometric errors for the IRAC (MIPS) bands, in order to reflect 
the systematic uncertainties of the instruments 
The main contributions to these 
uncertainties are due to the colour-dependence in the flat field and to 
the absolute calibration (see for example Lacy et al., 2005, and the IRAC 
and MIPS Data Handbook). 

For what concerns the optical photometry, in the SED fitting analysis
we adopted for each source the $U$, $B$, $V$, $R$, $I$ and $z'$ band from the Capak et
al. (2004) paper (Section \ref{optical}). We used the ACS photometry
only when this information is missing (see Table 3).




All the 19 sources in the final catalogue of spheroids with
$S_{24}>80 \mu$Jy and the 14 additional X-ray detections only are found 
within the spectroscopic Keck catalogue (Section \ref{spec}), and 
spectroscopic redshifts are also reported in Table 3.
Table 3 is splitted in three sections, one referring to the catalogue of 16 MIPS 
24$\mu$m sources with unambiguous optical identifications, the second to the 3 blended 
MIR sources, and the last one to the 14 spheroidal galaxies with X-ray detection only. 
For the potentially confused sources, $Spitzer$ (IRAC+MIPS) fluxes
have to be considered as upper limits only.

\section{DATA ANALYSIS}
In this Section we exploit the previously discussed multi-wavelength data 
to test the origin of the observed emissions and verify if these may be
ascribed to the standard passively evolving stellar populations, or instead 
require star-formation or AGN activities to take place in them.

To this end, we use the optical-IR Spectral Energy Distributions (SEDs) 
to check for the presence of hot dust emissions by a nuclear power-source 
(AGN) or emission by warm dust in star-forming regions. We will also look for 
the presence of emission lines the optical spectra and check for the ionization
level and line width to discriminate AGN activity.

Then optical to IR colours, X-ray to optical and X-ray to IR flux ratios are
used as diagnostics of the emission processes. In a few cases the very deep
X-ray data will provide unique diagnostic information.

\subsection{Analysis of the Spectral Energy Distributions for the IR-emitting Spheroids}
\label{seds}

\begin{figure*}
\begin{tabular}{cc}
\psfig{file=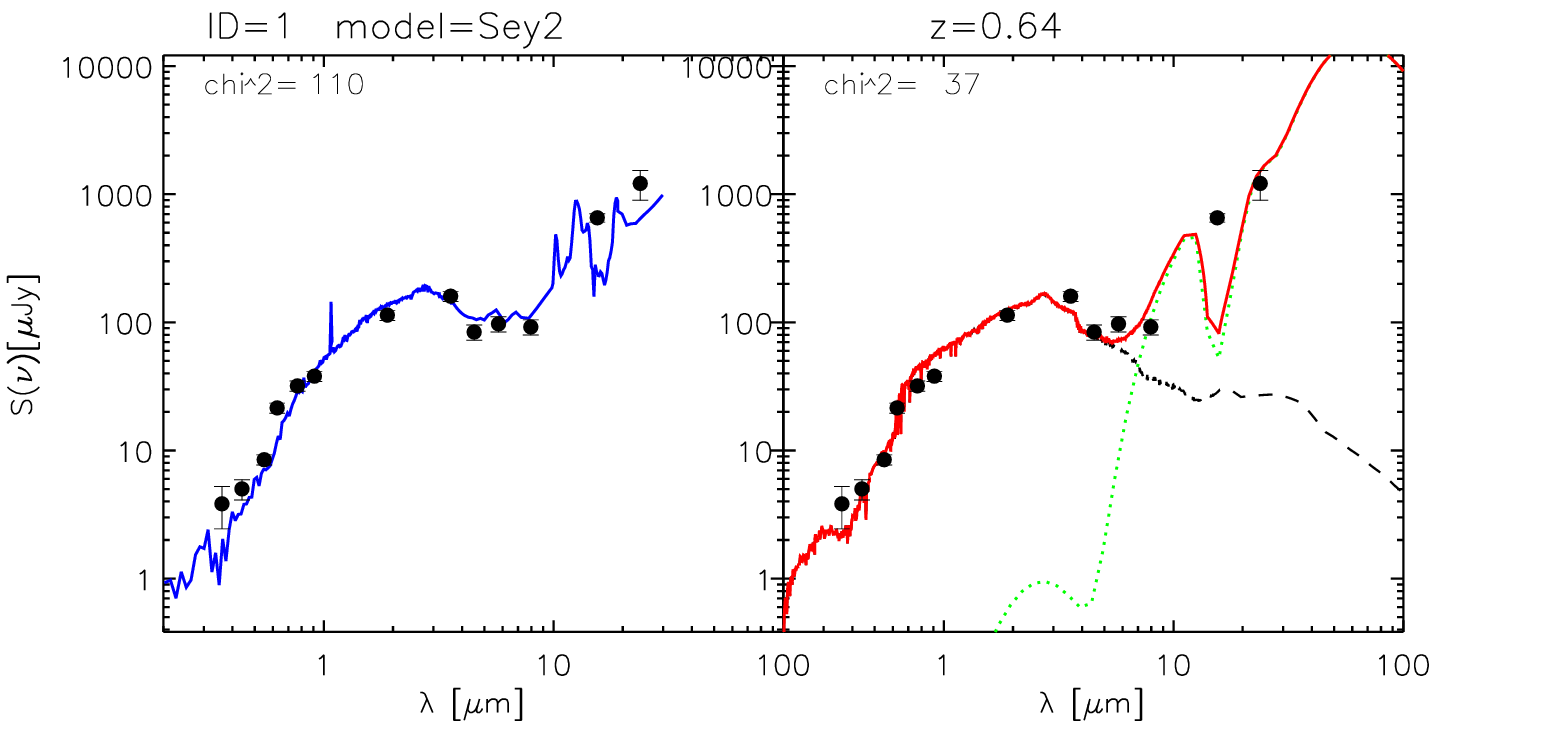,width=9cm,height=4cm}  & \psfig{file=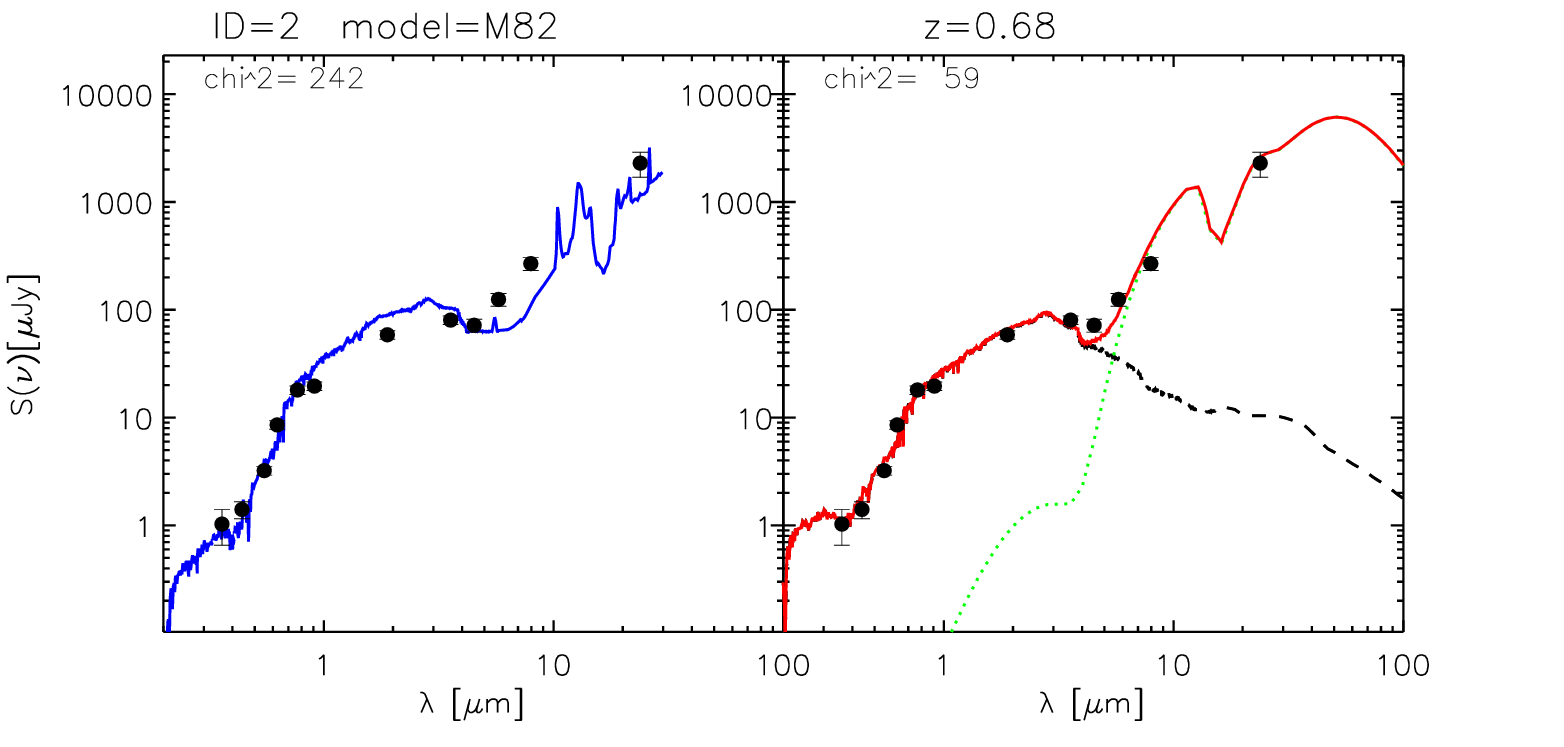,width=9cm,height=4cm}  \\
\psfig{file=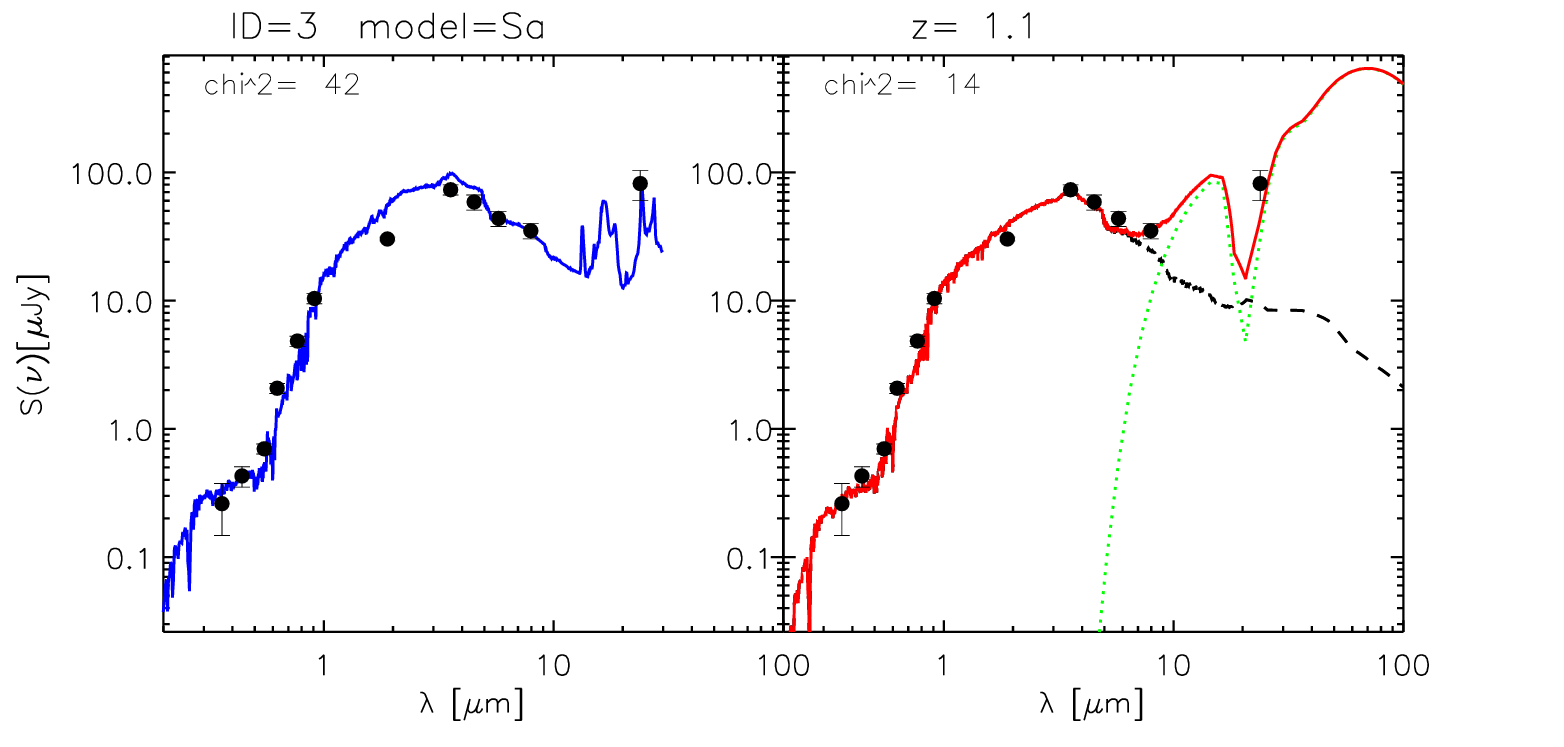,width=9cm,height=4cm}  & \psfig{file=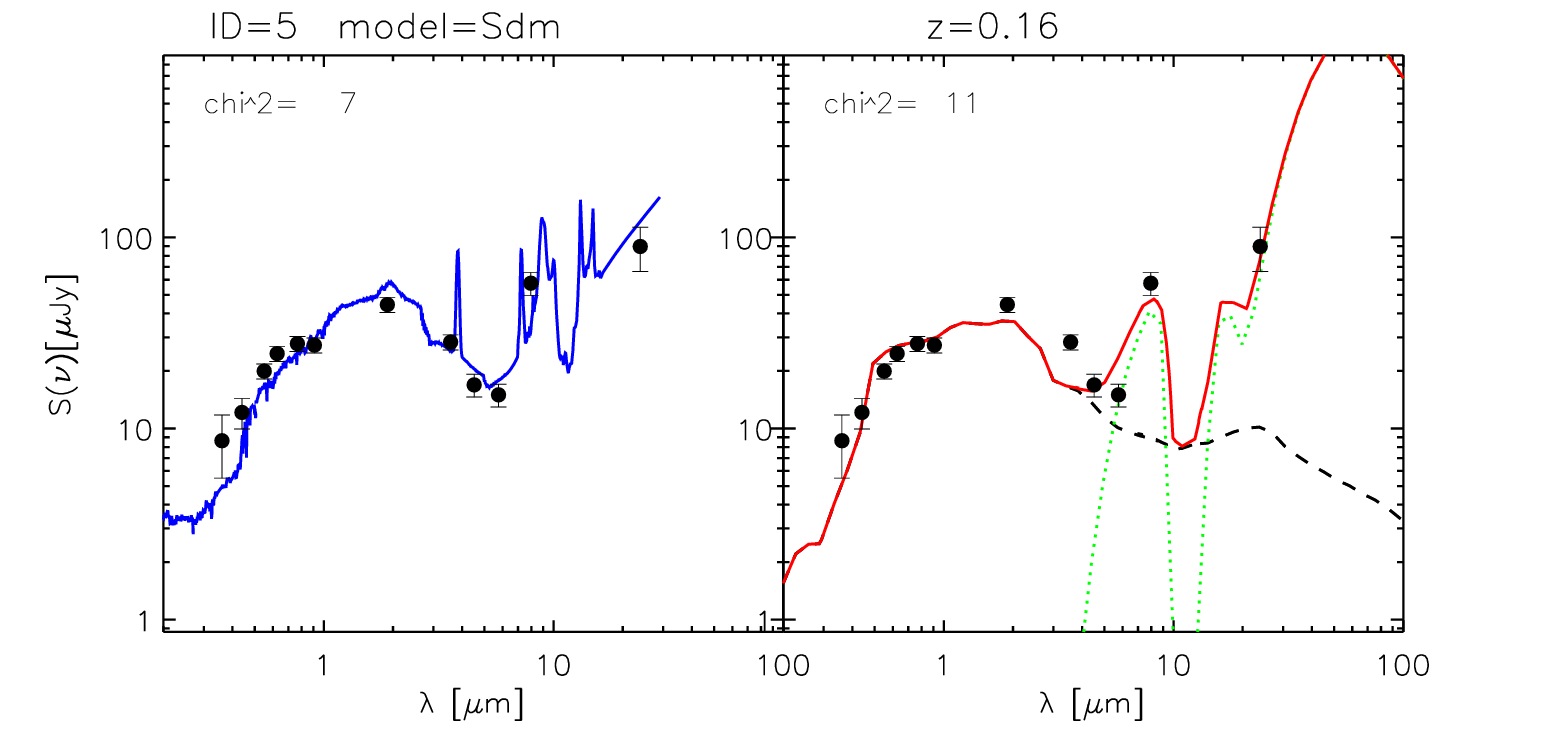,width=9cm,height=4cm}  \\
\psfig{file=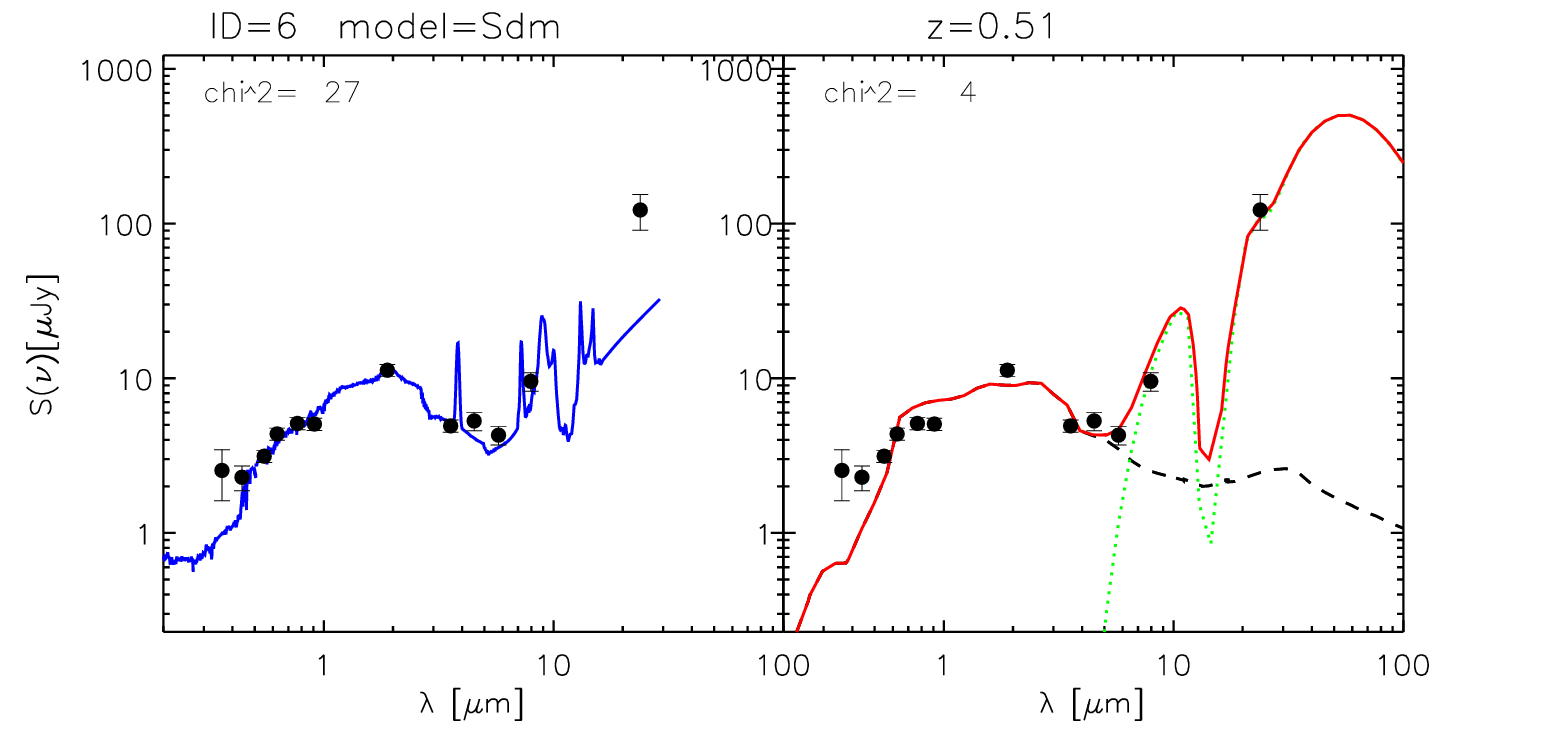,width=9cm,height=4cm}  & \psfig{file=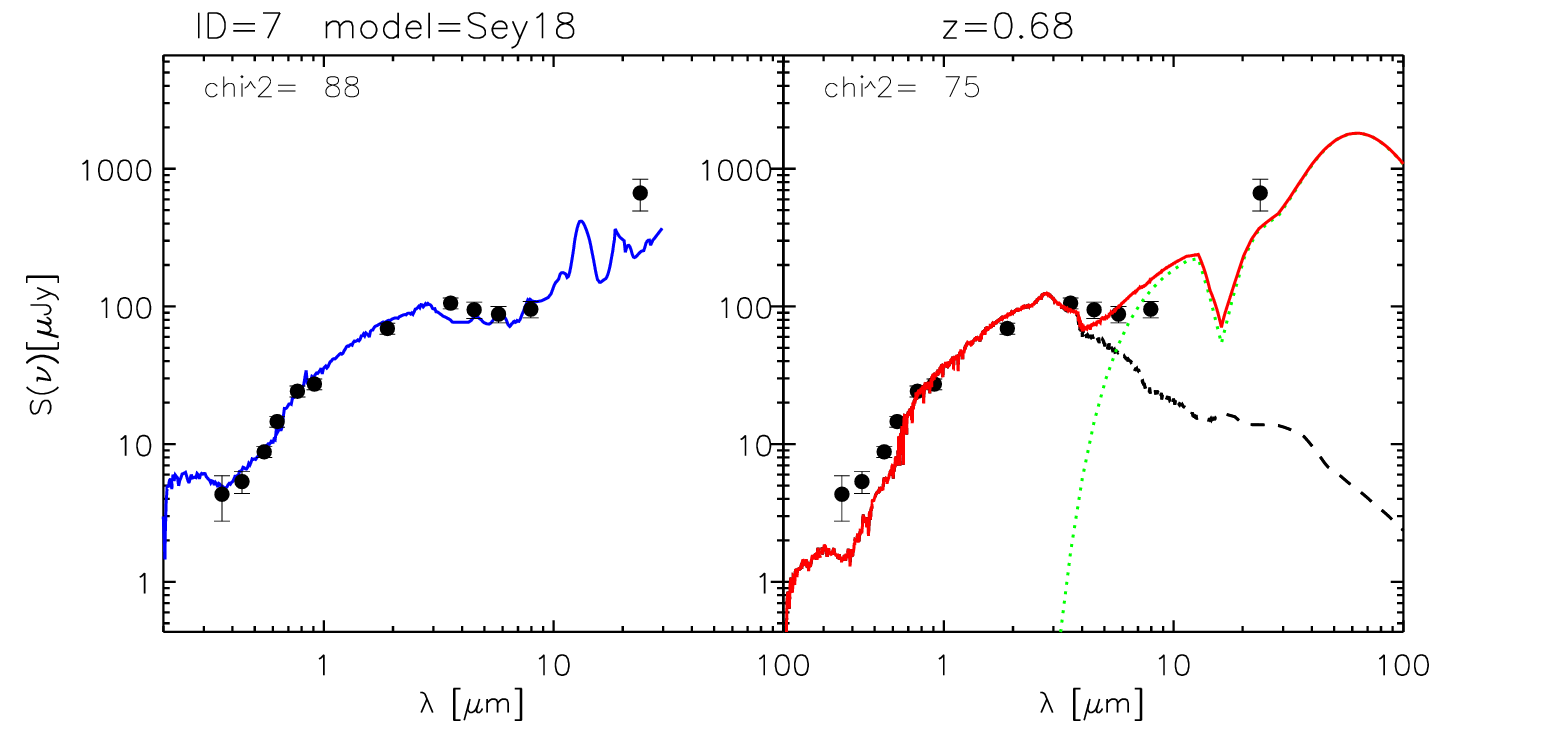,width=9cm,height=4cm}  \\
\psfig{file=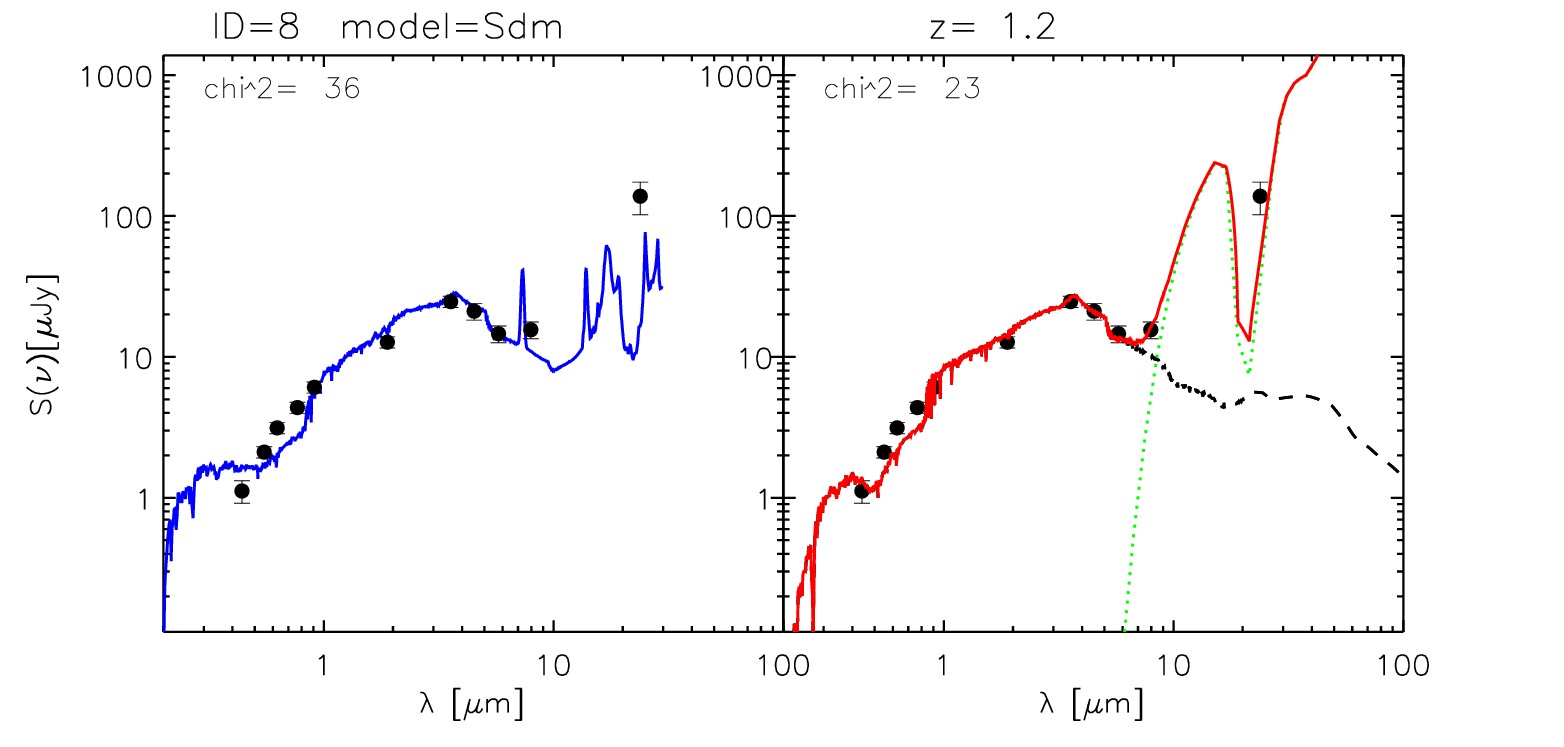,width=9cm,height=4cm}  & \psfig{file=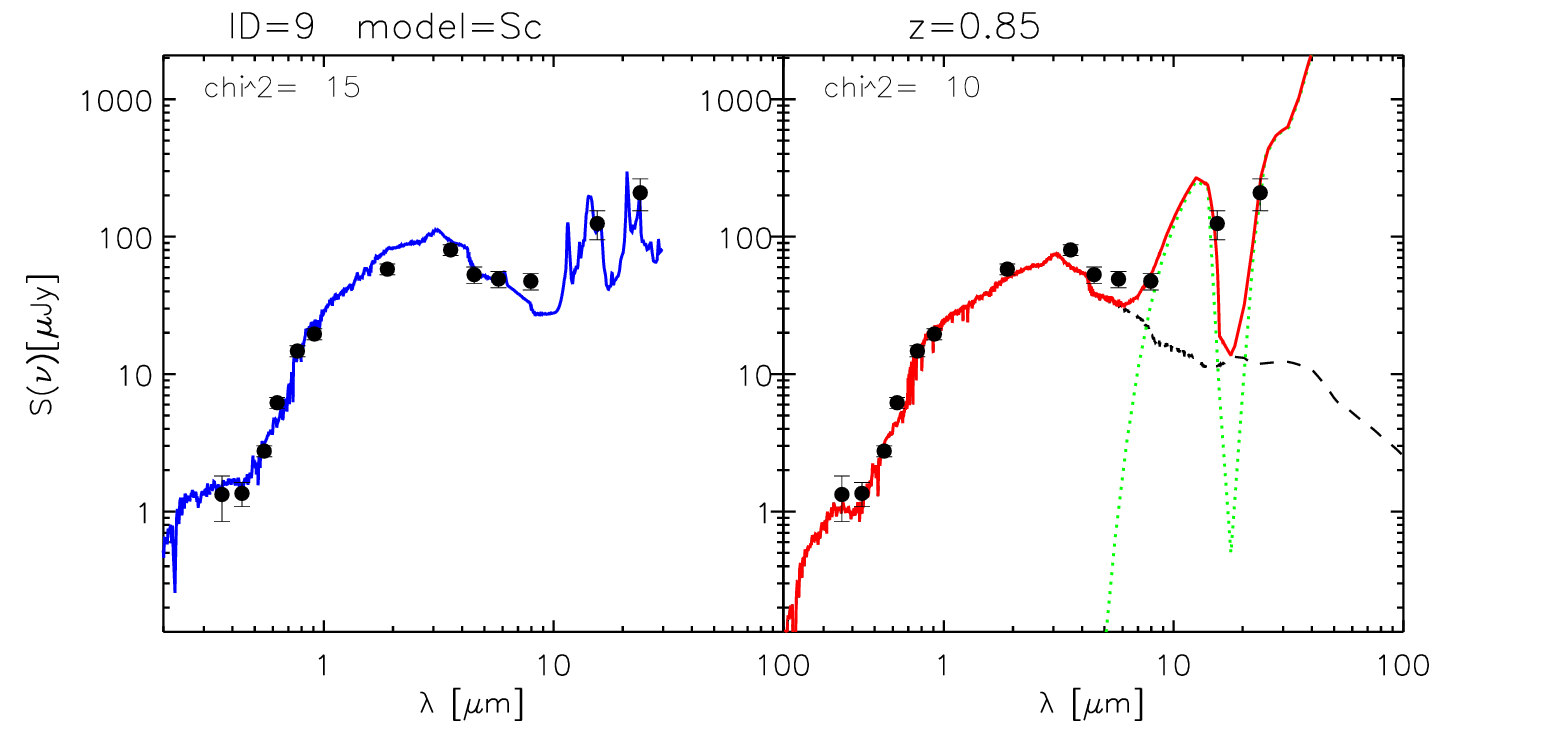,width=9cm,height=4cm} \\
\psfig{file=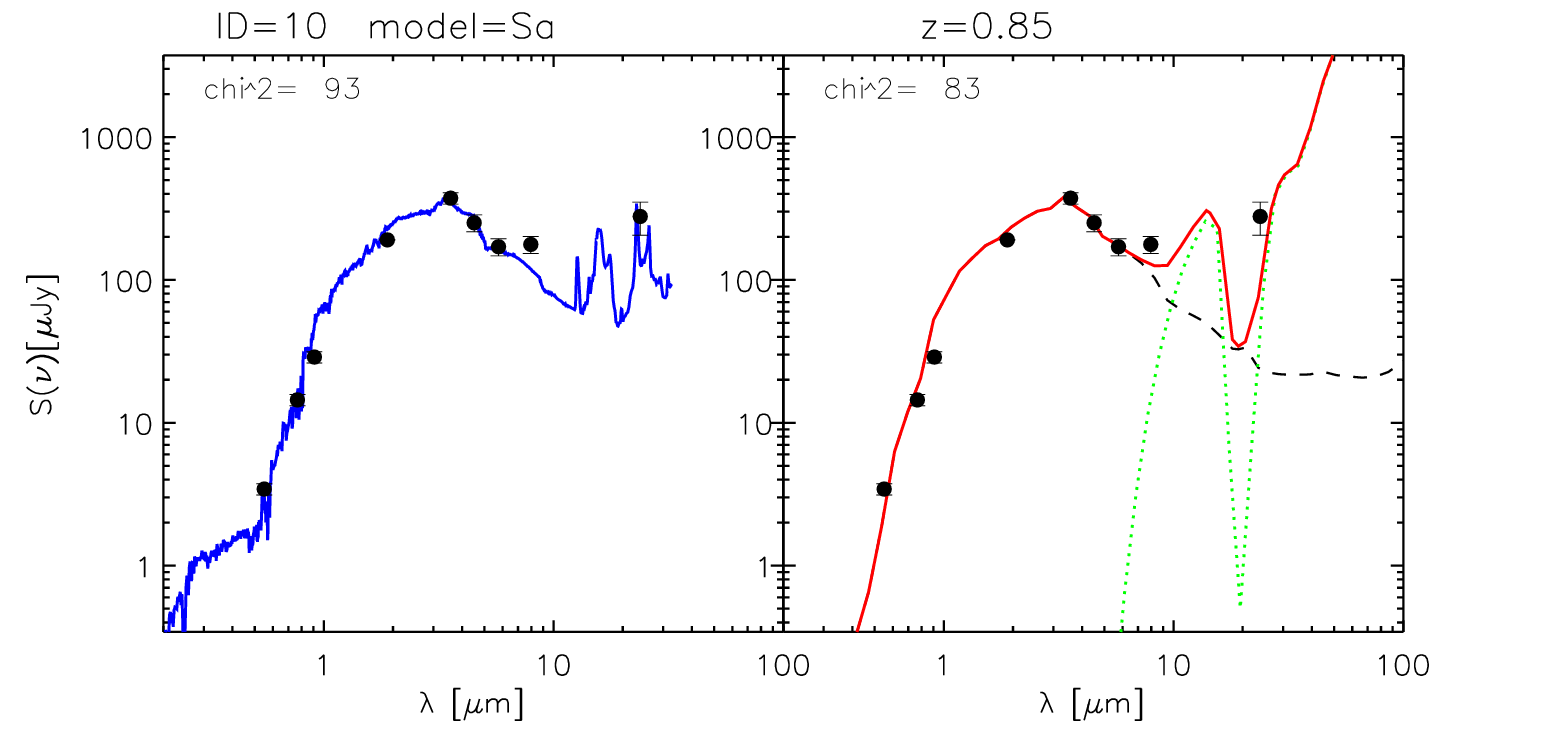,width=9cm,height=4cm} & \psfig{file=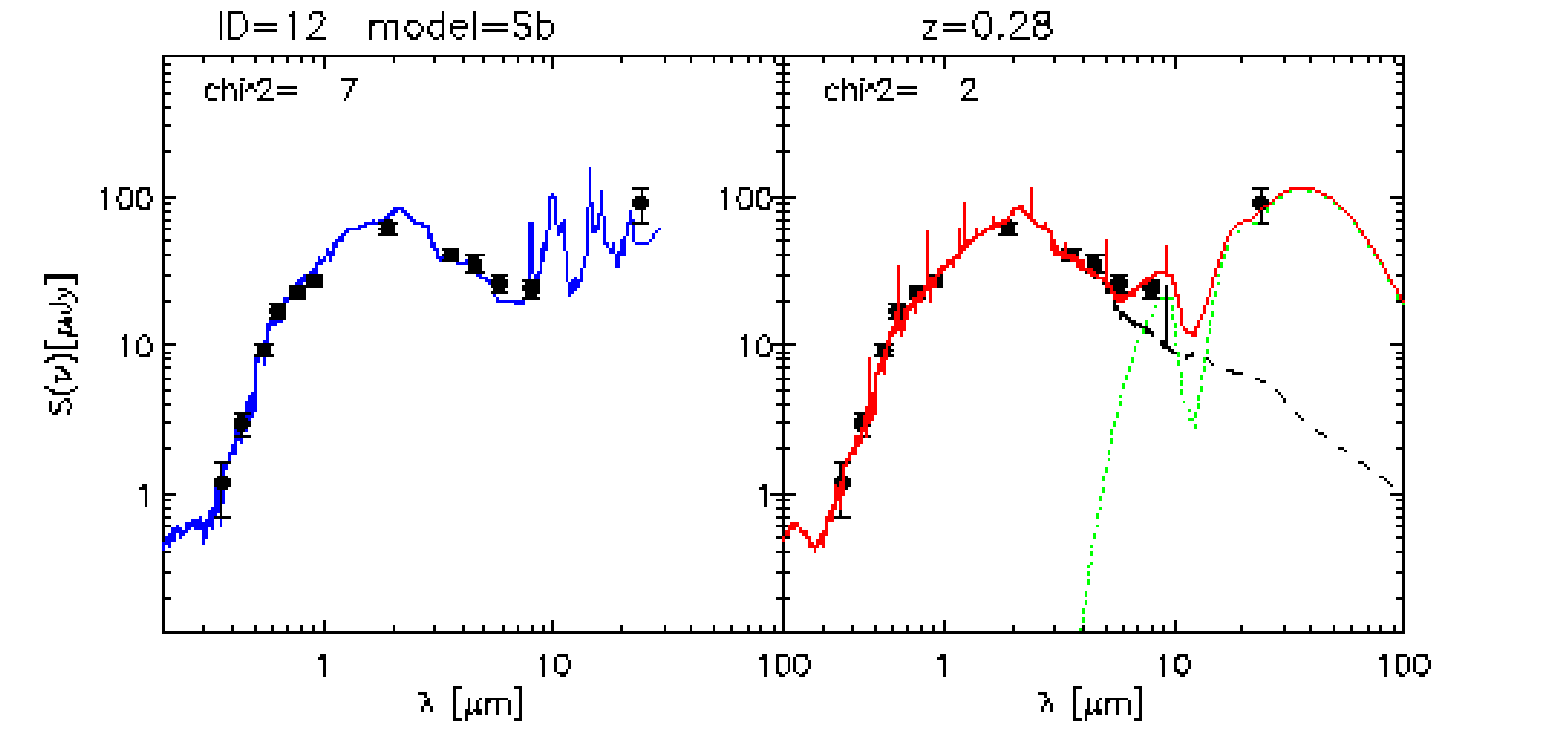,width=9cm,height=4cm} \\
\end{tabular}
\caption{Spectral Energy Distributions of spheroidal galaxies with detected 24 $\mu$m
flux. The left panels show a comparison of local template spectra with the observational SEDs.
The right panels illustrate two-component fits: an evolved stellar
population (dashed line) and dust obscured AGN emission (green
line). The sum of the two components is shown as the solid red line. 
 }
\label{fig_sed}
\end{figure*}

In order to understand the nature of dust emission by the IR-emitting
objects, we have assembled in Figures \ref{fig_sed} 
and \ref{fig_sed4_blend} the SEDs from UV to MIR for all of them. 
We have first attempted to interpret these observational SEDs of our spheroids 
galaxies using an automated fitting routine contained in the "Le Phare" code 
(available at {\tt http://www.lam.oamp.fr/arnouts/LE$\_$PHARE.html}).

Le PHARE (PHotometric Analysis for Redshift Estimations, Ilbert et
al. 2006) is a publicly available set of FORTRAN programs aiming at computing 
photometric redshifts through the best-fitting SED analysis. Le PHARE's structure is based on 
two parts:

$\bullet$ A preparation phase, composed of different programs in order 
to select theoretical SEDs, filters and build theoretical magnitudes. The different programs have 
been developed in order to easily get basic information relative to the filters 
($\lambda_{mean}$, AB-corrections ...) and the SEDs (k-correction
versus $z$, color-color diagrams for stars, quasars and galaxies ...).

$\bullet$ A running phase, based on the photometric redshift program 
itself. The program is based on a simple $\chi^2$ fitting method between the theoretical magnitudes 
and an observed photometric catalogue.

For our galaxies we already know the spectroscopic redshift, so we have 
fixed $z$ and have used Le PHARE only to obtain the best-fitting SED through a comparison 
with 21 templates of galaxies (including Ellipticals of different ages, lenticulars, 
Spirals and Starbursts) and AGNs (including low- and high-ionization QSOs, reddened QSO, Seyfert 
1, 1.8 and 2 and two kinds of ULIRGs) provided by M. Polletta et al. (in preparation).

Since the SED-fitting is based on the $\chi^2$ method, the most relevant 
photometric points are obviously those with smaller photometric errors (i.e. typically the 
optical/NIR ones).

From this first exploratory SED fitting we have inferred a rough spectral 
classification, and have looked for evidences of a MIR excess in the observed
24 $\mu$m fluxes compared with the best-fitting model spectra.

\begin{figure*}
\begin{tabular}{cc}
\psfig{file=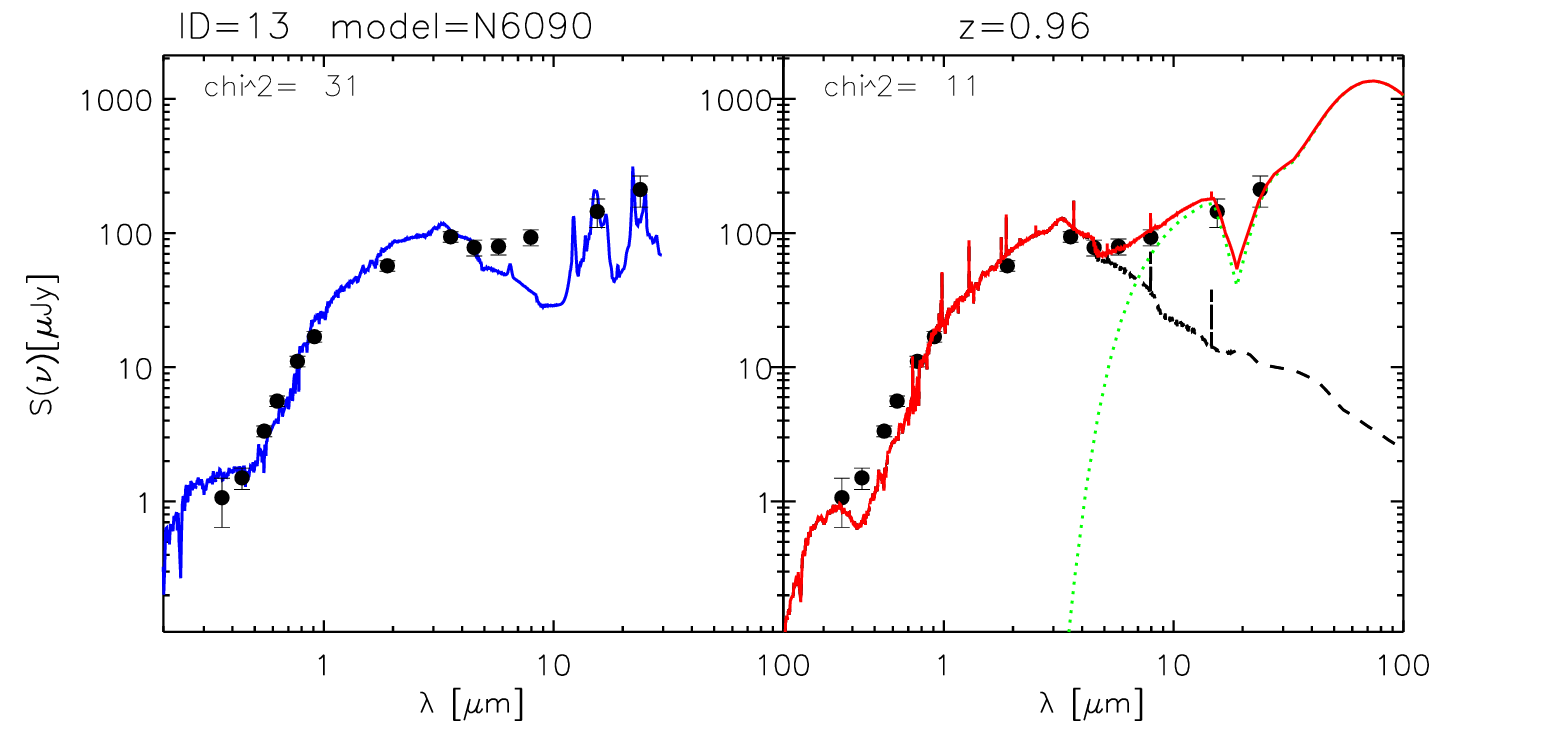,width=9cm,height=4cm} & \psfig{file=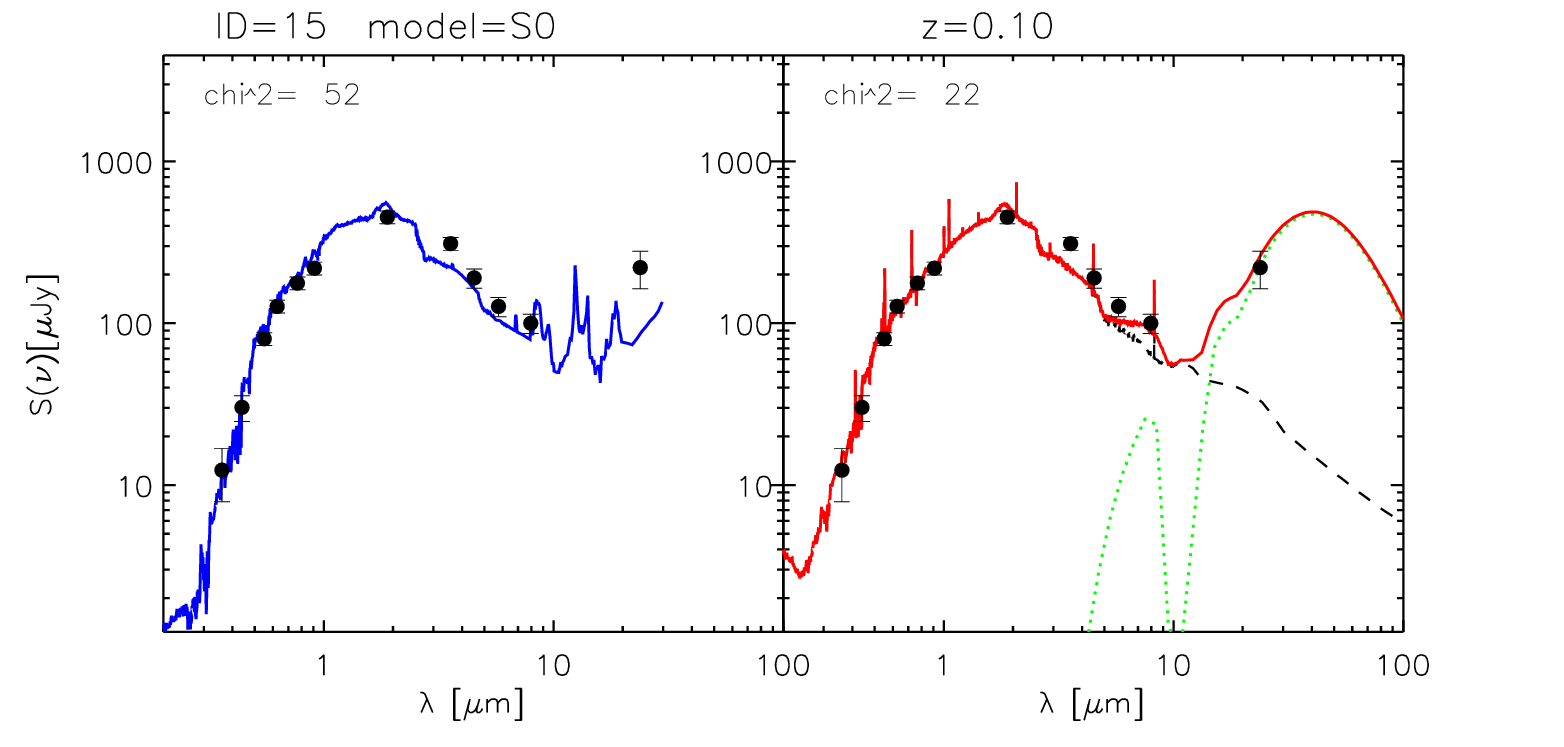,width=9cm,height=4cm} \\
\psfig{file=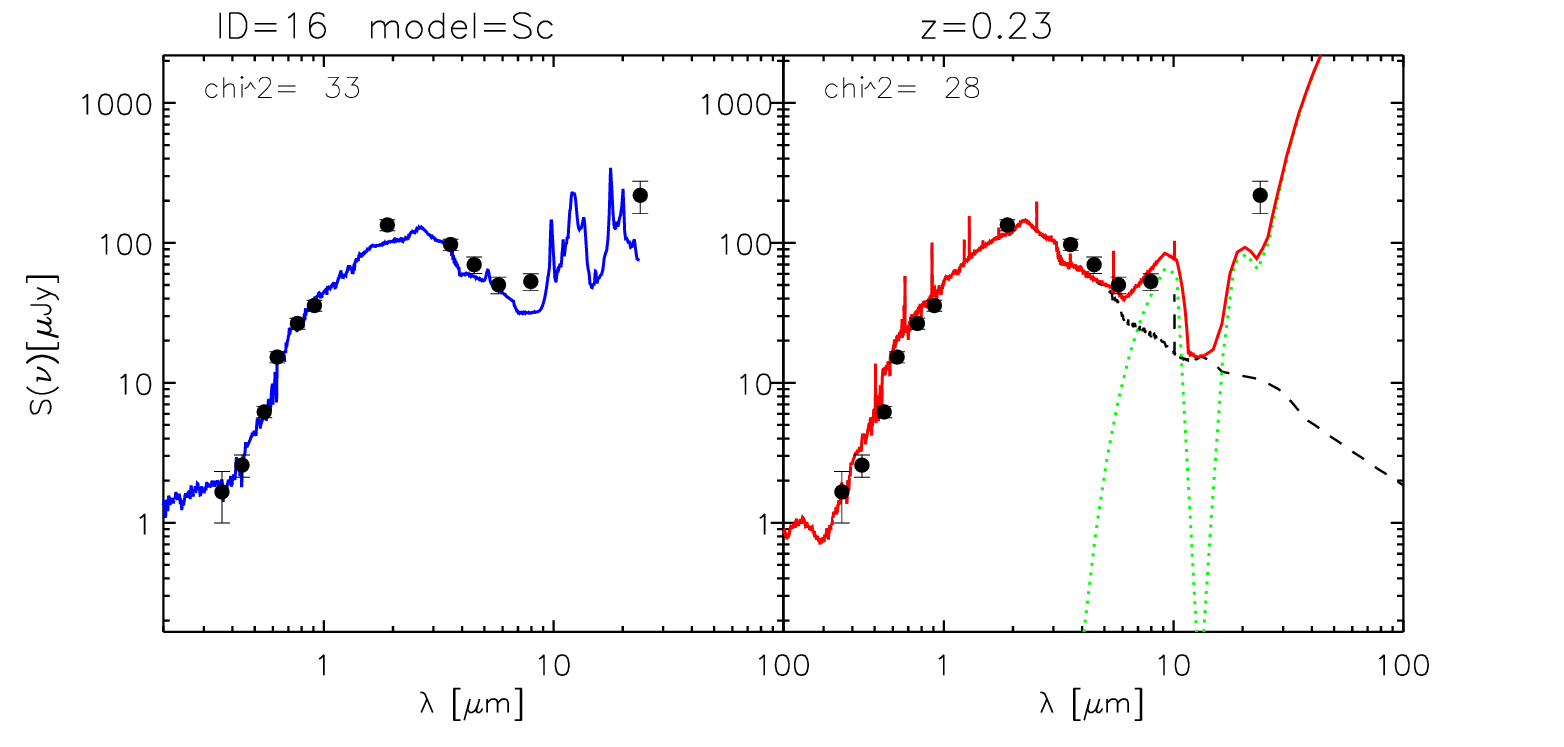,width=9cm,height=4cm}  & \psfig{file=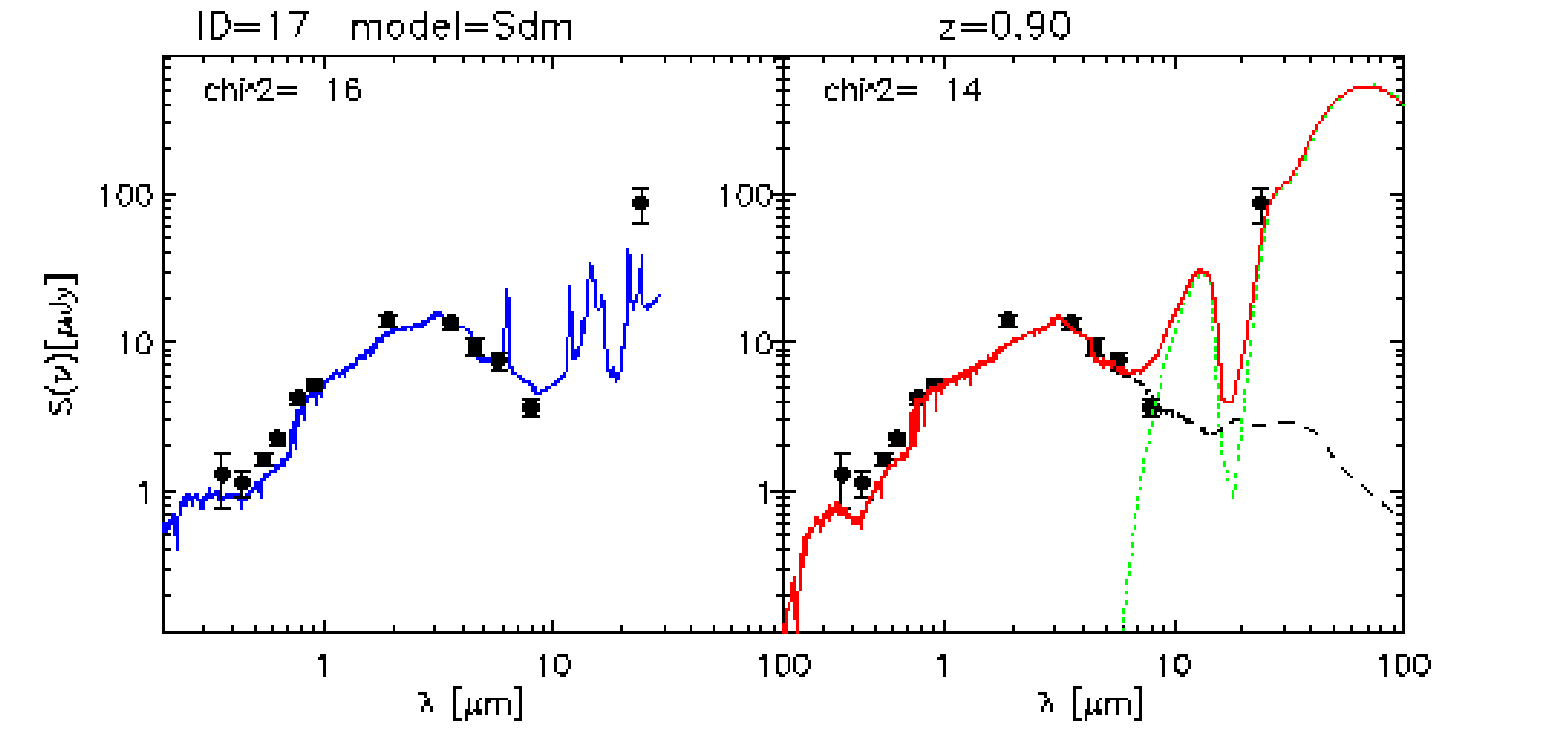,width=9cm,height=4cm}  \\
\psfig{file=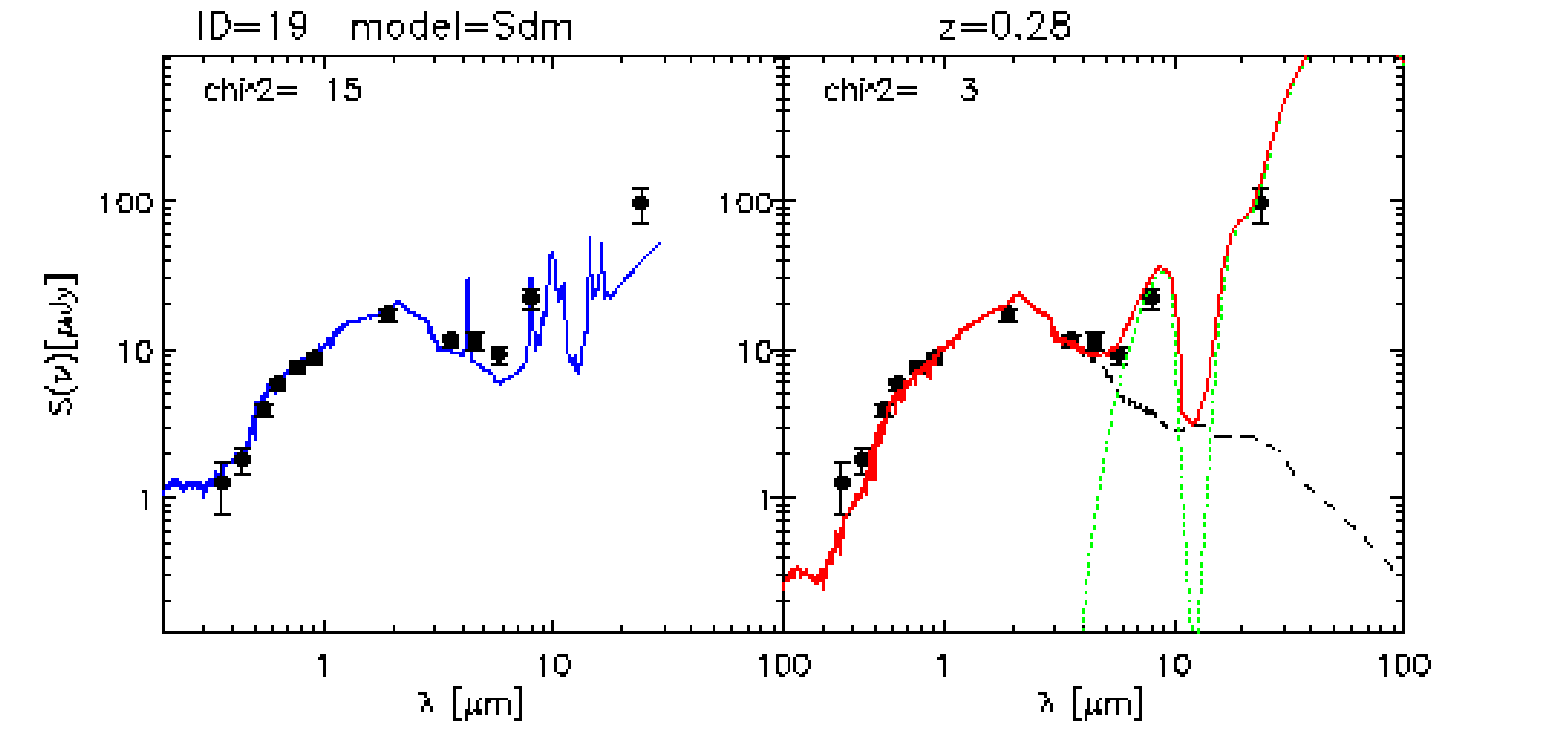,width=9cm,height=4cm}  & \psfig{file=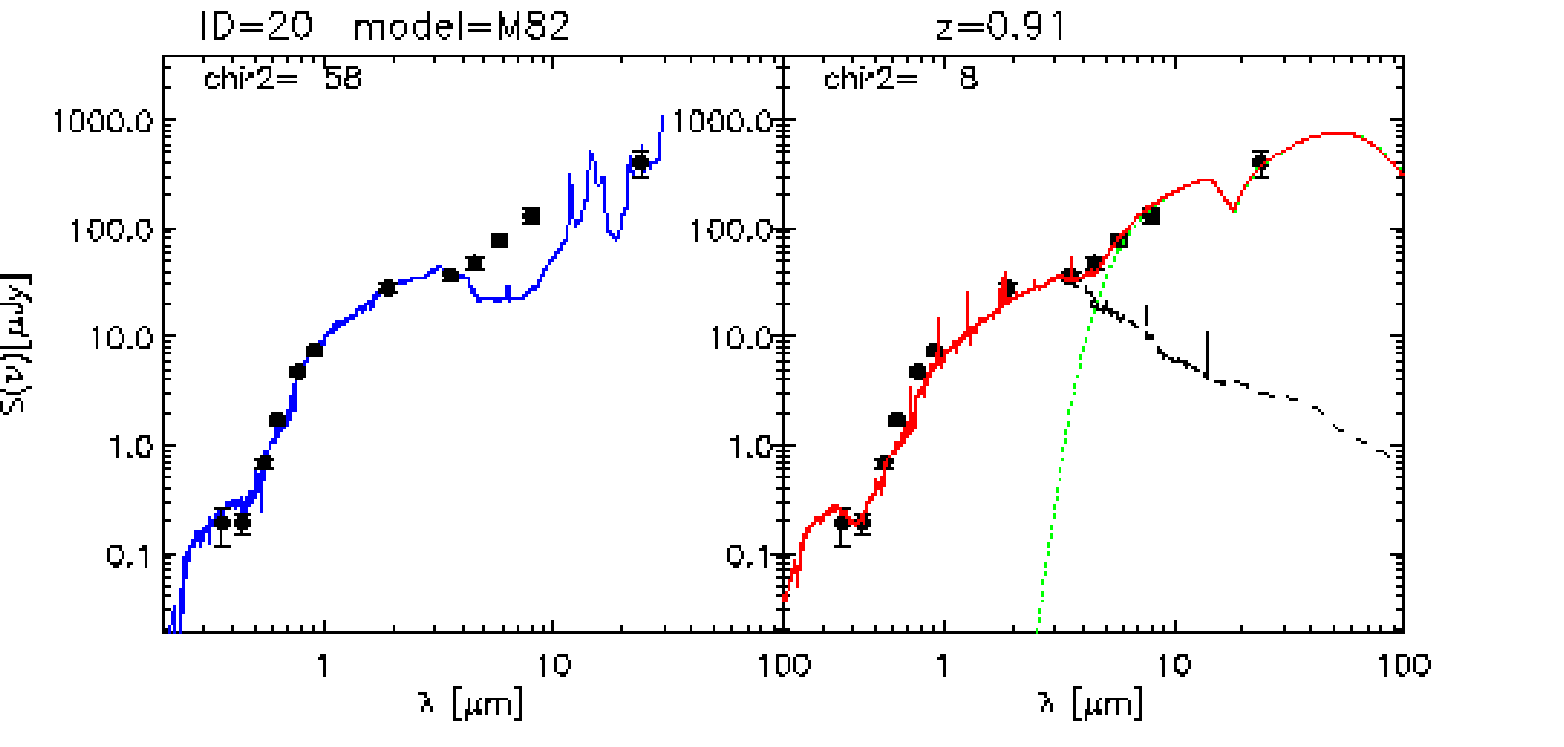,width=9cm,height=4cm}  \\
\end{tabular}
\begin{center}
Figure~\ref{fig_sed} (continued)
\end{center}

\end{figure*}

\begin{figure*}
\begin{tabular}{cc}
\psfig{file=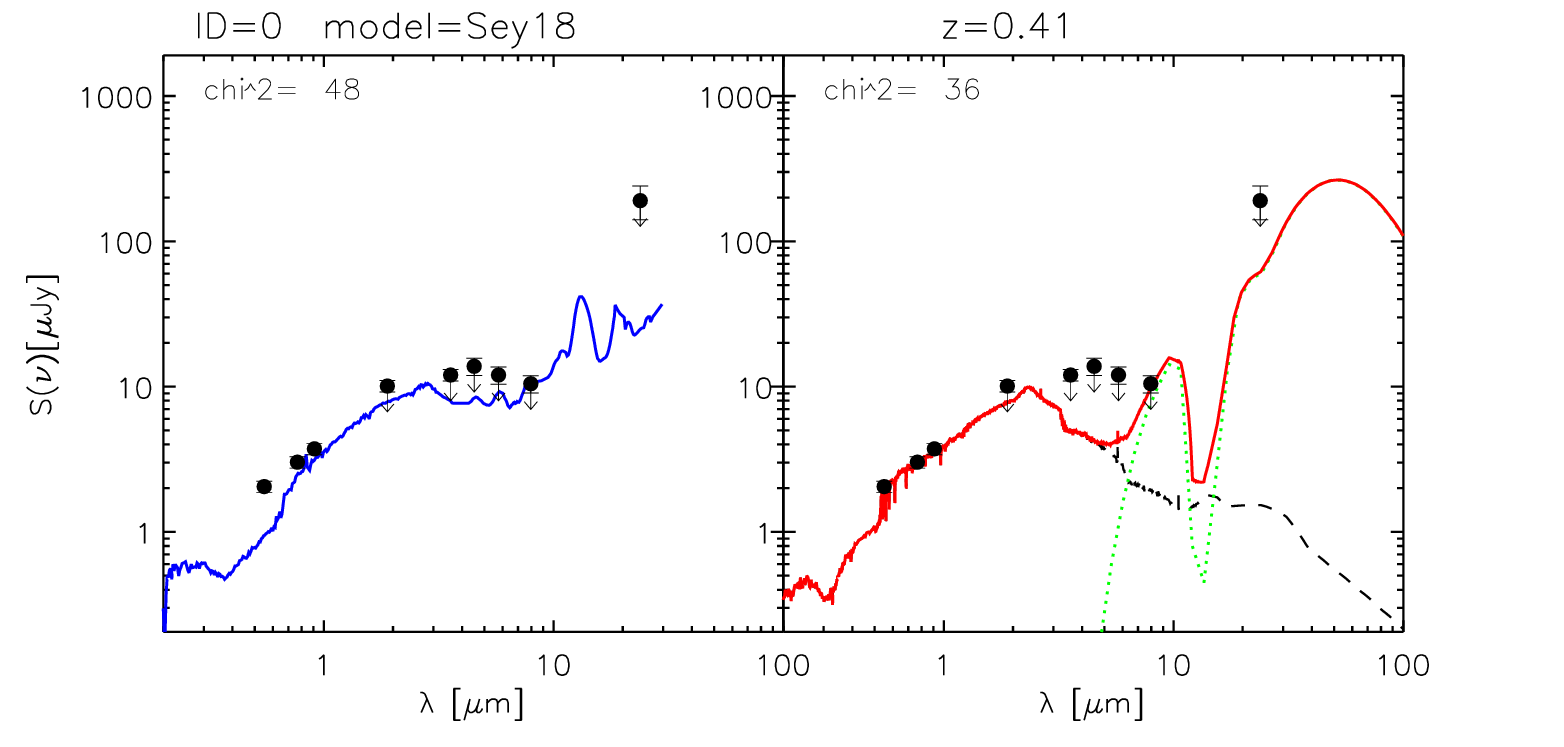,width=9cm,height=4cm}  & \psfig{file=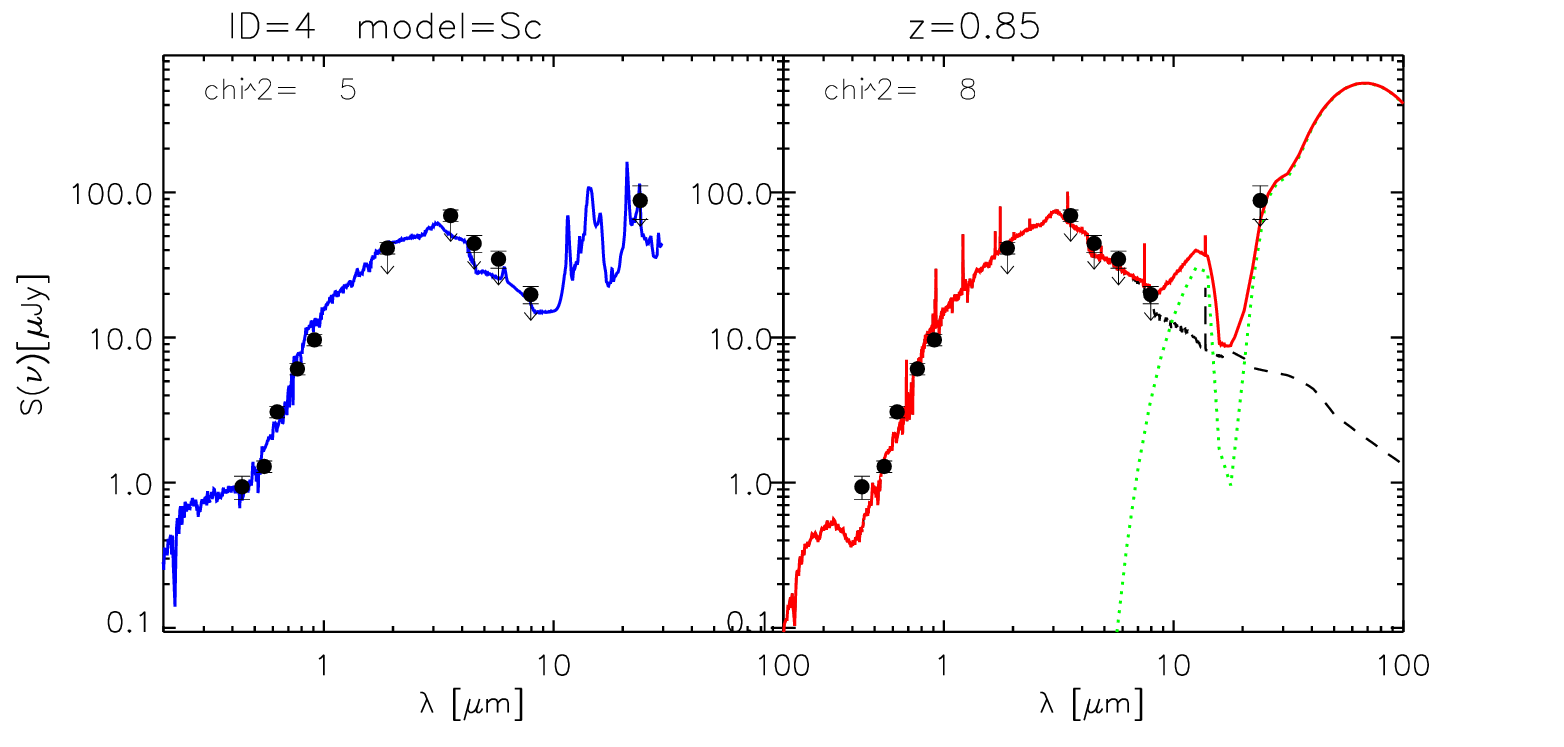,width=9cm,height=4cm}  \\
\psfig{file=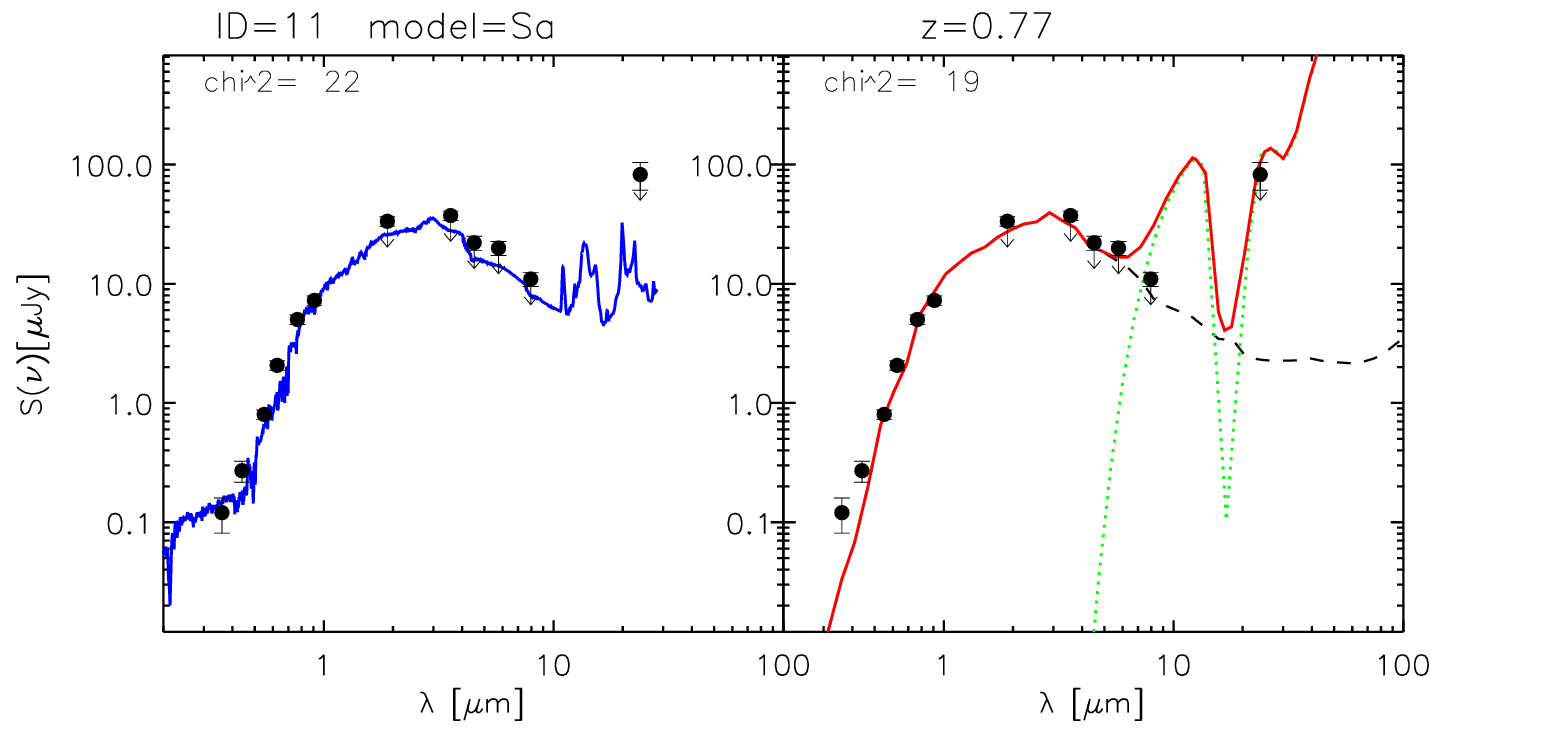,width=9cm,height=4cm}  & \\
\end{tabular}
\caption{Same as for Figure \ref{fig_sed} for the three MIR sources with multiple optical
associations (blended sources, see images in Figure \ref{blend}).
}
\label{fig_sed4_blend}
\end{figure*}

The left panels in Figures \ref{fig_sed} and \ref{fig_sed4_blend} provide a
comparison of our model fits based on local templates with the observational SEDs
for the spheroidal sample galaxies. For each source we report in the
plot a reference indicating the nature of the best-fit
model derived by the Le Phare code and the reduced $\chi^2$
corresponding to that solution. 
The large values of the $\chi^2$ statistics are mainly caused by the
inability of the adopted SED templates to reproduce the observed 24 $\mu$m
emission (but also the 5-8 $\mu$m fluxes of various sources). 
We note that the primary solution of this photometric analysis is
quite robust, within the uncertainties. In fact, only for a couple of sources the secondary
solutions derived by Le Phare are acceptable (in terms of minimization
of the  $\chi^2$) and imply a different spectral type (from an SED
dominated by star formation in favour of and SED dominated by an AGN,
or viceversally).
However, the main implication of this analysis is that
none of our sample spheroidal objects is reproduced by a typical
old and passive stellar population. On the contrary, only late-type
star-forming spirals or Seyfert galaxies seem to better match the
observed SEDs. This already implies that our selection naturally
favours the detection of a peculiar class of $active$
ellipticals (see discussion below). 

We see that in many cases (6 out of 19) the observed 24 $\mu$m fluxes
show a significant excess compared 
with the expectations based on the local spectral templates matched to
the optical-NIR data.  We discuss here possible interpretations of such MIR emission. 
Three different hypotheses are considered:
1) mass loss from late-type stars (Knapp et al. 1992), 2) IR emission by warm dust 
and PAH molecules related with star-forming regions (Madden et al. 1999;
Xilouris et al. 2004), 3) dust emission by circum-nuclear torii heated by an AGN 
(Knapp et al. 1999).

The first interpretation follows from the study by Knapp et al. (1992) of a sample
of nearby elliptical galaxies detected in the MIR (12 $\mu$m) by the {\em IRAS} surveys. 
The authors, after considering several possible origins for the 12 $\mu$m emission, 
conclude that for almost all the normal non-AGN ellipticals the MIR emission is likely
due to photospheric and circumstellar emission by evolved red giants stars in the galaxies.
The MIR (12 $\mu$m) and NIR (2.2 $\mu$m) fluxes of the nearby ellipticals are 
compared by Knapp et al. (1992) to those of evolved red giants, showing the existence
of a large population of mass loosing stars. MIR and NIR fluxes are then used by them 
to derive rates of stellar mass loss.

\begin{figure}
\centerline{\psfig{figure=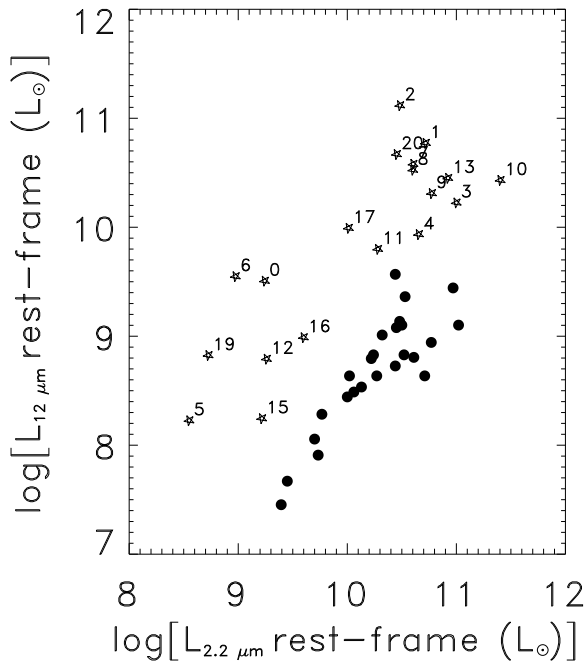,width=14.5cm}}
\caption{Relation of the 12 $\mu$m versus 2.2 $\mu$m rest-frame luminosities for MIR emitting
spheroidal galaxies. The GOODS-N early-type galaxy colours (open circles) are compared to
those of nearby elliptical galaxies studied by Knapp et al. (1992; filled circles), and
show a large excess in the 12 $\mu$m luminosity.} 
\label{fig_l12}
\end{figure}

To test the hypothesis that the MIR emission from early-type galaxies in our sample 
has the same origin of the nearby ellipticals, we have compared the NIR and MIR 
properties of the two samples.  Since our galaxies are at a median redshift 
$<z> \simeq 0.68$, from our observed 3.6 and 24 $\mu$m fluxes we have computed 
rest-frame luminosities at 2.2 and 12 $\mu$m with suitable K-corrections
(that turned out to be rather insensitive to the adopted template spectrum),
and then compared them in Figure \ref{fig_l12} with intrinsic 12 and 2.2 $\mu$m
luminosities of local objects.

The values of the MIR to NIR luminosity ratios for our early-type galaxies 
(open stars) turn out to be about a factor of ten higher on average than for the 
Knapp et al. nearby ellipticals (filled circles).   
This would imply a fraction of 12 $\mu$m luminosity produced by mass loss 
(and the relative mass loss rate) in our galaxies far too high if compared
with the values found for local galaxies with MIR excess. We conclude that, if some
contribution to the observed MIR emission by mass loss from evolved giant stars is 
present in our spheroidal galaxies, it is certainly not the dominant energy source 
at these wavelengths.

The presence of diffuse dust emission, responsible for the observed MIR excess in the 
spectra of early-type galaxies, has been considered by several authors 
(i.e. Madden et al. 1999; Xilouris et al. 2004).
To test this hypothesis for our galaxies, we have considered the results of the automatic 
SED fitting described in the previous section and shown in the left panels of Figs. 
\ref{fig_sed} and \ref{fig_sed4_blend}.
In 8 galaxies out of 19, the optical-to-MIR SEDs can easily be fitted with normal galaxy 
templates and residual star-formation. 
In at least 6 further cases there is instead evidence that the rest-frame 
MIR emission is significantly in excess of any reasonable level that might be explained
by diffuse dust in star-forming regions.

Our third considered hypothesis is therefore that the MIR excess observed in the GOODS-N 
early-type galaxies is contributed by the presence of AGN circum-nuclear dust re-radiation.
This interpretation is suggested by the shapes of the observed IR SEDs, which exclude the 
alternative possibility of direct non-thermal quasar unobscured emission.


We have then tried to fit the observational data with a combination 
of two different spectral components:
an evolved stellar population to reproduce the optical and NIR spectrum,
plus an obscured AGN emission component to explain the MIR data. 
The synthetic SEDs of the old component, including a mix of passively evolving stellar 
populations and younger stars, have been modelled with a population-synthesis code 
reported by Berta et al. (2004), based on the Padova stellar
isochrones. 
The synthetic spectra are taken from a set of templates used in Fritz et
al. (2006, submitted). We found that sources ID5 and
ID6 are reproduced with a typical evolved old stellar population of 1
Gyrs, while ID 10 and ID 5 with an older population of 4 Gyrs.
Other six sources (IDs 0, 1, 3, 8, 9 and 17) are better represented
by a galaxy during a post-starburst phase with a second main episode of star formation
--forming $\sim 10\%$ of the total stellar mass-- in the range $\sim 10^8-3\times10^9$ years.
The remaining objects (IDs 2, 4, 7, 12, 13, 15, 16, 19 and 20) are
instead fitted with a main--impulsive burst
and a continuous star formation which is truncated at $2\times 10^7$
years.
The best-fit solutions for the stellar component are also used to
provide a measure of the barionic mass of each source (a Salpeter IMF
in the range 0.15-120 $M_{\odot}$ has been considered). 
The stellar masses are reported in the last column of Table 1, with a
median value of $log(M)\sim 10.7\ M_{\odot}$.

The AGN IR component is modelled as the emission by a dusty torus heated by 
a central AGN, as detailed  below.
The right-hand panels in Figs. \ref{fig_sed} and \ref{fig_sed4_blend} 
illustrate the two component (old population $+$ dusty torus) solutions.

Our adopted torus model has been discussed in Fritz et al. (2005), to which we
refer for a more detailed explanation. 

The model reproduces the AGN spectrum as 
a combination of a nuclear non-thermal spectrum plus a dusty torus emission.
A flared disc geometry, i.e. two concentric spheres with the polar cones
removed, is adopted for the latter.  The torus size is defined by the angular 
opening angle and by the radius of the outer sphere, while the inner radius 
is set by the sublimation temperature of graphite and silicate grains 
under the influence of the strong nuclear radiation field.

The computation of the thermal and scattering emissions
throughout the torus is performed by numerically solving the 
radiative transfer equation with an iterative procedure.

As discussed in Fritz et al., the model provides excellent fits to
the observed SEDs of a wide variety of active nuclei and quasars.

\begin{center}
\begin{table}
\begin{minipage}{200mm} 
  \caption{Object Classification}
 \scriptsize
\begin{tabular}{c c c c c c}
\hline
\hline
  ID &   X-ID  &      z  & optical spectral & Overall SED    & Mass\\  
     &         &         & classification   & classification & ($M_{\odot}$)\\  
\hline
 0 &   ~   & 0.410  & ELG                   &     AGN      & 10.88 \\        
 1 & X-067 & 0.638  & Early+[OII]           &     SB       & 12.15 \\              
 2 & X-082 & 0.679  & AGN2                  &     AGN      & 11.21 \\              
 3 & X-110 & 1.141  &                       &     AGN      & 11.34 \\              
 4 & X-113 & 0.845  & LINER                 &     AGN      & 11.32\\                
 5 &   ~   & 0.156  & ELG                   &     SB       & 10.05 \\        
 6 &   ~   & 0.512  & LINER                 &     AGN?     & 9.50 \\         
 7 & X-115 & 0.680  & AGN2                  &     AGN      & 10.65 \\         
 8 & X-149 & 1.223  & ELG                   &     SB/AGN?  & 10.78 \\                
 9 & X-160 & 0.848  & Starforming           &     AGN      & 11.50 \\           
 10 & X-169 & 0.845  & Early-type           &     AGN?     & 11.10 \\         
 11 &   ~   & 0.766  & Early+[OII]          &     SB       & 11.02 \\         
 12 &   ~   & 0.277  & Early-type           &     SB       & 9.76 \\         
 13 & X-240 & 0.961  & AGN2                 &     AGN      & 12.50 \\         
 15 & X-383 & 0.105  &                      &     SB       & 10.25\\         
 16 & X-388 & 0.231  & Early-type?          &     SB       & 10.40 \\             
 17  &   ~  & 0.899  & Starforming          &     SB       & 10.20 \\         
 19  &  ~   & 0.278  & Starforming          &     SB       & 9.44\\         
 20  &   ~  &0.911   & Early+faint[OII]     &     AGN      & 10.74\\         
\hline
 1x  &  X-414   ~  &0.798   & Early                &     SB        & 10.90\\         
 2x  &  X-230   ~  &1.013   & Early-type, post-SB  &     SB?       & 11.51\\         
 3x  &  X-294   ~  &0.474   &                      &     AGN       & 10.14\\         
 4x  &  X-249   ~  &0.475   &                      &     SB        & 10.56\\         
 5x  &  X-286   ~  &0.954   & Early                &     AGN       & 11.40\\         
 6x  &  X-257   ~  &0.089   & Early-type           &    Inactive   &9.79\\         
 7x  &  X-274   ~  &0.321   & Early-type           &     Faint AGN?& 10.74\\         
 8x  &  X-354   ~  &0.569   &                      &     SB?       & 10.44\\         
 9x  &  X-241   ~  &0.850   & Early-type           &     AGN       & 10.66\\         
10x  &  X-210   ~  &0.848   &                      &     SB?       & 11.26\\         
11x  &  X-212   ~  &0.943   & Early-type           &     AGN       & 10.90\\         
12x  &  X-194   ~  &0.556   & Early+faint[OII] &     AGN   & 10.34    \\         
13x  &  X-131   ~  &0.632   & Starforming          &     SB        & 10.20\\         
14x  &  X-114   ~  &0.534   & Early-type           &     SB        & 10.66\\         
\hline								   
\hline 								   
\end{tabular}							   
\end{minipage}
\label{tabb}
\end{table}
\end{center}

\begin{figure*}
\psfig{file=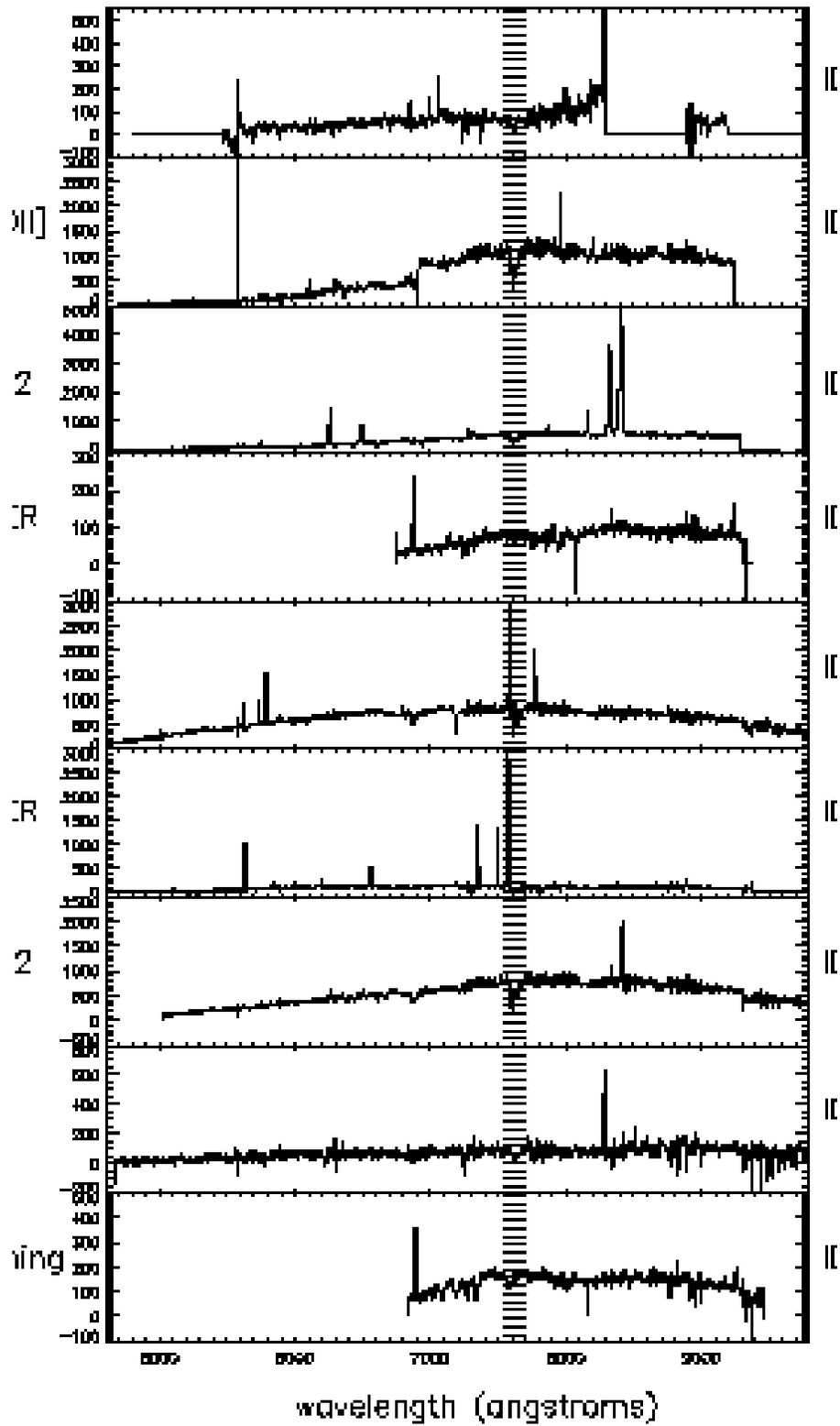,width=12cm}
\caption{Observed optical spectra of spheroidal galaxies with excess MIR
emission. The ID numbers are those reported in the top panel of
Table 3. The dashed region indicates an absorption feature
of the sky.}
\label{specz1}
\end{figure*}

\begin{figure*}
\psfig{file=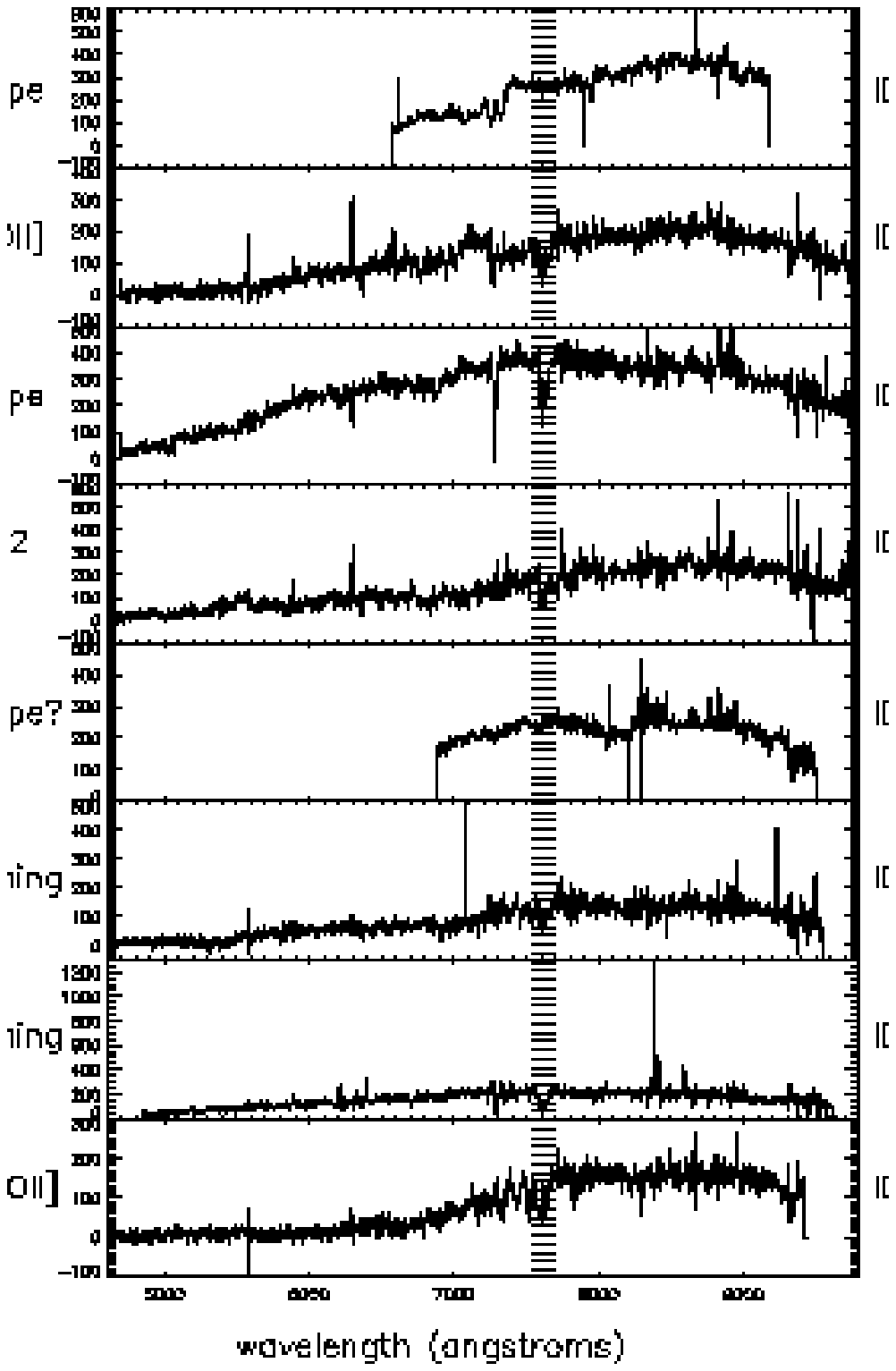,width=12cm}
\begin{center}
Figure~\ref{specz1} (continued)
\end{center}
\end{figure*}

In almost all cases, the combination of an evolved stellar population fitting 
the optical and NIR spectrum, and an obscured AGN emission component to 
explain the MIR spectrum, always provides much better fits to the data (as
can be argued from the comparison of the $\chi^2$ statistics in
the left and right panel of Figures \ref{fig_sed} and \ref{fig_sed4_blend}).
As anticipated, in 6 cases this solution including an AGN is required by the IR data.
(Obviously, we did not attempt to constrain the torus physical parameters with
the few available data-points).

In order to have a complete census of the physical properties of the
spheroidal population, we checked the observed optical-to-IR SEDs of the
$z$-band selected galaxies lacking a detectable mid-IR or X-ray
counterpart (135 out of the initial 168). For these sources we built
a multiwavelength photometric data-set similar to that described for
the main sample of this work, covering the spectral range 0.3-8.0 $\mu$m. 
The vast majority of these objects show SEDs consistent with that of
a passively evolving stellar population, typical of quiescent
elliptical galaxies. We found that only few sources ($\sim$6\%)
present some indication of ``activity'' in the observed SEDs, simply
because the photometric data points at wavelengths larger than 5
$\mu$m show an underlying rising spectra.  
However, a more detailed work on this photometric data-set would be
required to constraint the properties of the full morphological spheroidal sample.

\begin{figure*}
\psfig{file=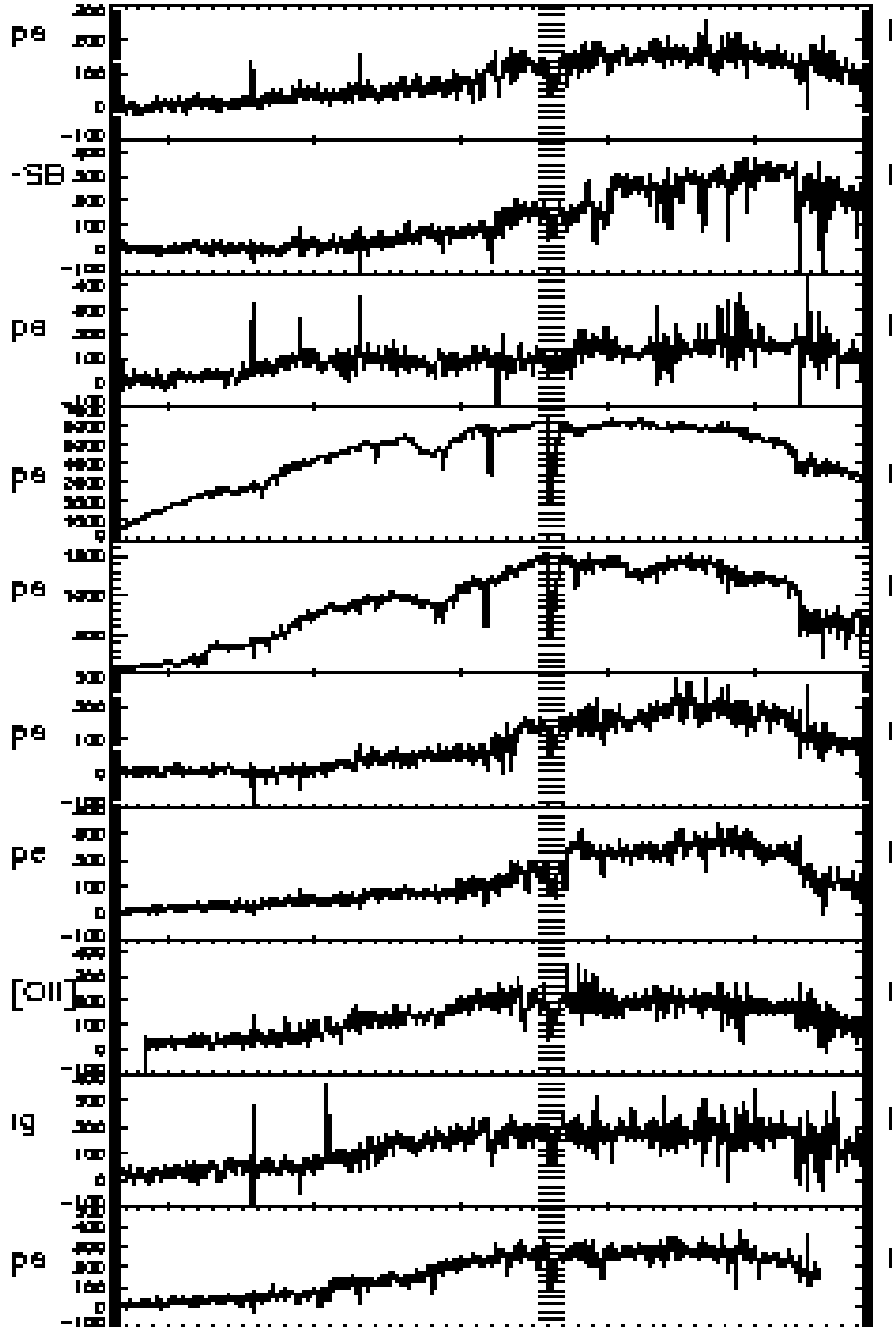,width=12cm,height=18cm}
\caption{Optical spectra of spheroidal galaxies with excess X-ray but 
no MIR emission. The ID numbers are those reported in the bottom panel of
Table 3.}
\label{specz}
\end{figure*}

\subsection{Optical Spectroscopy}
\label{opt_spec}

The optical spectra available for our high-redshift spheroidal
galaxies are reported in Figures \ref{specz1} and \ref{specz}.
The emission lines were
analysed only in terms of the equivalent widths and flux ratios of contiguous 
lines, because the optical spectra retrieved from the WEB were not flux calibrated.
Objects with a red continuum and faint (or absent) emission lines
($EW< 5\AA$) are classified as passive early-type galaxies. When two or more emission
lines were measured, we applied the usual diagnostic diagrams 
(Veilleux \& Osterbrock, 1987) in order to classify the galaxies according to
the star-forming emission line, Seyfert-2 or LINER categories. 
In few cases we could not disentangle the Seyfert-2s/LINERs classes 
due to the limited observed spectral range.
Sources with more than one emission line were classified as Starforming,
while when only a single emission line is visible, we 
classify the object as an Emission Line Galaxy (ELG).

\begin{figure*}
\psfig{file=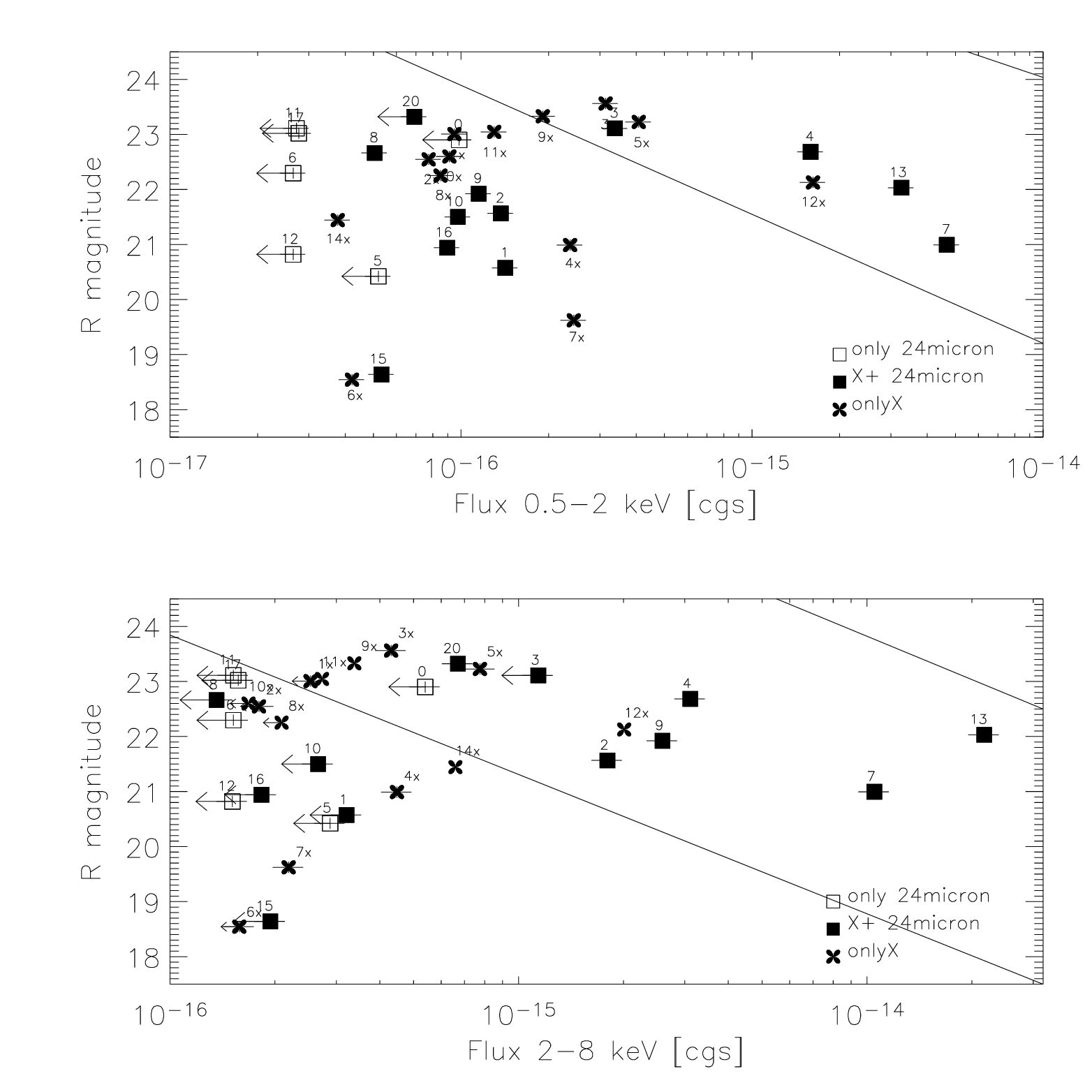,width=14cm,height=14cm,angle=0}
\caption{Top panel: the R--band magnitude versus the 0.5-2 keV 
X-ray flux for the 12 {\it Spitzer}--MIPS ellipticals with 
X--ray counterparts (filled squares), those without X--ray counterparts
(open squares), and the spheroids detected only in the X-rays but not in 
the MIR (crosses).  The area between the two continuous lines represents 
the locus typically occupied by known AGNs, and corresponding to $log(X/O)=0\pm1$.
Bottom panel: the same plot against the 2--8 keV X-ray band.}
\label{fxR}
\end{figure*}


\begin{figure}
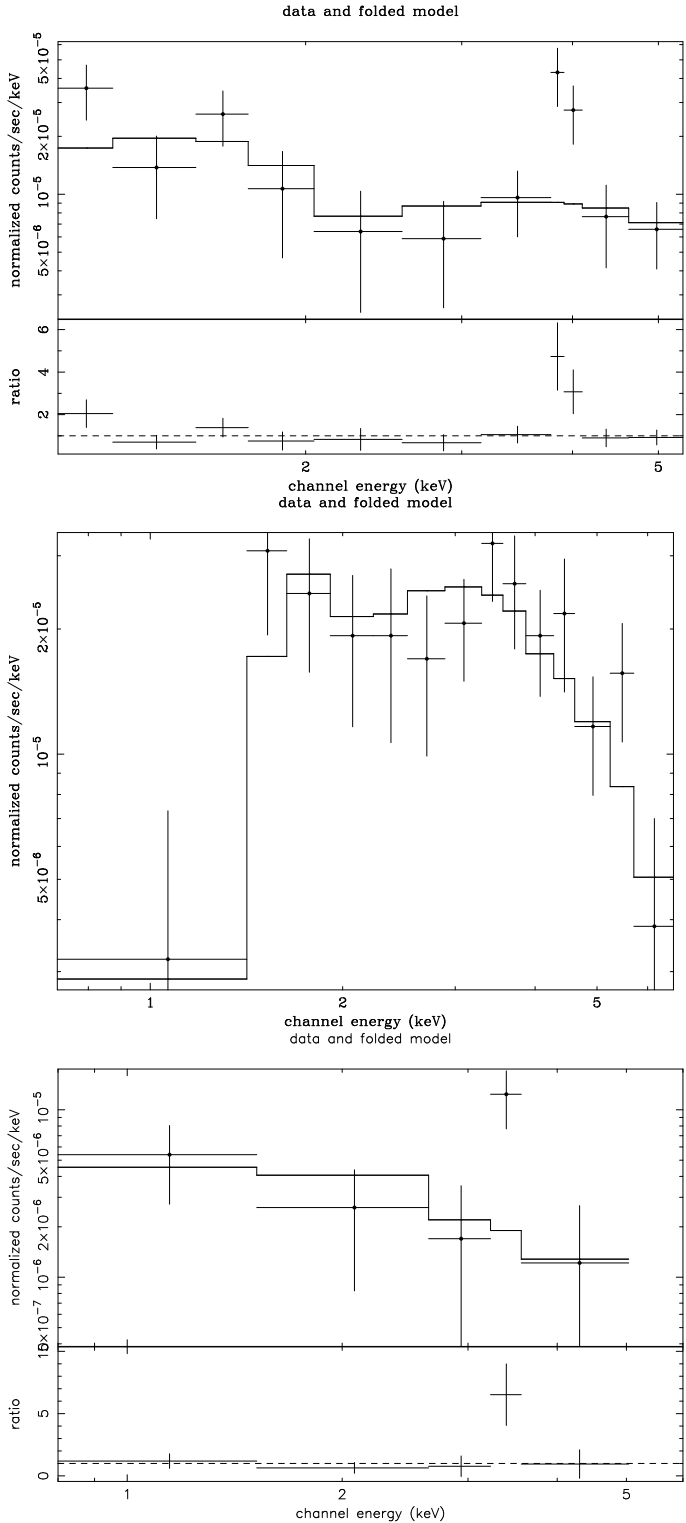

\psfig{file=82.ps,width=6.5cm,angle=-90}
\psfig{file=source9_gamma1.8.ps,width=7.1cm,angle=-90}
\psfig{file=468.ps,width=6.5cm,angle=-90}
\caption{
The X--ray spectra of sources 2,9 and 20 (from top to bottom)/
  An absorbed power law model with both spectral index and 
 absorption free to vary has been adopted for source \#9. 
 The flat X--ray spectrum of source \#2 is well fitted by a single power law model 
 (see Table 2).  The low counting statistic of source \#20 does not
  allow to perform a proper spectral fit. A power law with slope fixed at $\Gamma=1.8$ 
 is  adopted to show excess emission around the energy expected 
 for a redshifted iron K$\alpha$ line. 
}
\label{240}
\end{figure}

Optical spectra for our sub-sample of galaxies detected only in the 
X-ray bands are reported in Figure\ref{specz}, and have been analysed
based on similar procedures.

Since the original parent sample is morphologically selected spheroidal
galaxies, we might be surprised by the high incidence of strong emission lines in
the optical spectra (67\% of the objects show at least one line with $EW>9\AA$),
considering the usual fairly good agreement between the morphological and
spectroscopic classifications. However, this fraction of optical emission 
line spectra is not untypical if referred to the original parent sample: a 10\% 
of emission-line objects among high-z morphological ellipticals was also found 
in the K20/GOODS sample by Cassata et al. (2005).
Moreover, it is well known that the fraction of spheroids with
emission lines increases with redshift, with a typical value of
$\sim$20\% in the redshift range $0.5<z<1$
(Brinchmann et al. 1998; Treu et al. 2002; Treu et al. 2005).
In particular, $\sim$25\% of the spheroidal galaxies in the Treu et
al. sample, when limited to the same morphological classes considered
in this work (class=0,1), present OII emission lines ($\sim$10\%
with $EW> 5\AA$).
This might suggest that the fraction of active spheroids based only on
optical indicators is consistent with that determined with the
combination of X-ray and MIR selections.

\subsection{X-Ray Properties}
\label{xray}

We consider here the X-ray emission properties of spheroidal 
galaxies. A first indication comes from Figure \ref{fxR}, plotting the soft and hard 
X--ray fluxes versus the R--band magnitudes. 
The first point to be remarked is that the optical and X-ray fluxes appear
largely unrelated. 

The values of the X--ray--to--optical flux ratio ($X/O=\log{\frac{f_X}{f_R}}=\log{f_X}+C+\frac{R}{2.5}$,
with C=5.5) can then be used to obtain a 
preliminary classification of the X--ray emission components (e.g. Maccaccaro 
et al. 1988): various classes of X--ray emitters are
characterized by different values of  X--ray--to--optical flux 
ratio.
Optically and X--ray selected AGN have, on average,
relatively high X/O values (--1$<X/O<$1), while infrared emitting starforming 
galaxies show lower X/O ratios ($-2<X/O<-1$; Alexander et al. 2001;
Franceschini et al. 2003; Ranalli et al. 2003), and even lower values 
($X/O<$ --2) are typical for X--ray quiescent normal galaxies, whose X--ray
flux is due to the integrated emission of compact sources and/or hot interstellar 
plasma (Hornschemeier et al. 2001). 

This classification based on the X--ray--to--optical flux ratio should be considered 
with caution and is somewhat dependent on the adopted energy range. 
For this reason we consider separately the 0.5--2 keV and 2--8 keV X--ray 
fluxes in the figure. The six sources detected in the hard band (filled squares 
in bottom Figure \ref{fxR}) are well within the AGN locus. Their soft X--ray 
fluxes are lower than the 2--8 keV ones and the observed 0.5--10
keV luminosities are within the Seyfert range (10$^{42-44}$ erg/s).

Three of them (IDs \#2, \#9, \#20) fall outside the AGN region ($-1< X/O < 1$)
in the soft band diagram, and, according to the classification
based on the X-ray-to-optical flux ratio, they would be classified
as star-forming galaxies. However, the AGN locus boundaries in $X/O$
diagrams should not be considered as strict limits. 
From an inspection of the two panels of Figure 9, it is clear that the three sources have
hard spectrum'
This hypothesis is corroborated by the results of the spectral
analysis.

Five out of the six hard X--ray detected sources (2, 4, 7, 9, 13) have enough
counts ($>$ 100) to allow us to perform a moderate--quality spectral
analysis.  Source and background spectra, response matrices and
effective areas have been extracted from the merged events file using
the standard {\tt CIAO} tools developed to properly weight responses
and effective area files for multiple extraction regions ({\tt
mkwarf} and {\tt mkwrmf}; see Civano et al. 2005 for a detailed
description).  Source spectra were re-binned in order to have enough
counts per bin (15--20) to apply $\chi^2$ statistic, and were fitted
with {\tt XSPEC 11.3} (Arnaud 1996).  
Power--law model spectra with intrinsic absorption at the 
source redshift have been adopted.  
The results of spectral fitting are reported in Table 2.
Errors are at the 90 per cent confidence level for
one interesting parameter ($\Delta\chi^2 = 2.71$). 
The best fit $N_H$ values and the corresponding unabsorbed 
0.5--8 keV luminosities (10$^{43}$--10$^{44}$) are typical of moderately 
obscured Compton--thin Seyfert galaxies (Risaliti et al. 2002). 

Despite the relatively low counting statistics, the
X--ray spectrum of ID \#9 (Figure \ref{240}) does clearly
require a rather large column of cold gas ($log(N_H) > 23~ cm^{-2}$).
Source \#2 has a very flat spectrum and an emission line at the redshift expected from
the iron $K_{\alpha}$ with a large (1 keV) EW is evident from
the residuals of a power law fit (Figure \ref{240}).
A similar model provides a good description of the ID \#20's spectrum, though
with lower significance due to the poor counting statistics.
In both cases, Compton thick absorption is the most likely
explanation of the X--ray properties.
Assuming that the observed luminosity is entirely due to reflection by
thick matter with an albedo of 2\% (Comastri 2004), the intrinsic
luminosity would be of the order of about $10^{44}$ erg/s for source
\#2 and about $10^{43}$ erg/s for source \#20.

 \begin{table} 
 
\begin{minipage}{200mm} 
  \caption{X-ray spectral analysis parameters}
 \scriptsize
\begin{tabular}{ c c c c c c c c c c c c c c}
\hline
\hline

AID &	id  &z &$\Gamma$    & NH$\times$10$^{22}$  & chi$^2$/dof &L$_X$(0.5-10 keV)\\
     	  										  
82\footnote{Source number 82 has been fitted with the Cash statistic.}  & 2   &0.679    & 0.40$\pm$0.33      &     &~              $^a$	  &0.1$\times$10$^{43}$ \\
113 & 4   &0.84     &1.56$^{+0.12}_{-0.20}$	& $<$0.6   & 39/31   &  1.0$\times$10$^{43}$ 	\\
    &	  &	    &1.8		    & 0.7$\pm$0.3	& 41/32   & 1.0$\times$10$^{43}$\\
115 & 7   &0.68     &1.57$\pm$0.07	& 0.49$^{+0.22}_{-0.13}$  & 63/62    &2.1$\times$10$^{43}$\\
    &	  &	   &1.8		 & 0.84$\pm$0.13    & 73/63& 2.1$\times$10$^{43}$\\
160 & 9   &0.848    & 1.10$^{+0.80}_{-0.57}$	& 9.8$^{+8.2}_{-5.05}$        & 6/11	 & 8.7$\times$10$^{42}$\\
    &	  &	     &1.8		  & 15.1$^{+6.2}_{-3.9}$     & 8/12 &9.7$\times$10$^{42}$\\
240 &  13 &0.961   &1.59$\pm$0.09	& 4.8$^{+0.6}_{-0.5}$    &105/96  &1.3$\times$10$^{44}$\\
    &	  &	    &1.8		& 5.59$\pm$0.36   & 112/97& 1.6$\times$10$^{44}$\\
\hline
\hline 
\end{tabular}
\end{minipage}
\label{tabb}

\end{table}


Finally source \#3 is also classified as an AGN on the basis of the 
observed X/O in the soft band (it is undetected in the hard band) and the
relatively high ($\simeq 7 \times 10^{42}$ erg/s) X--ray luminosity.
Therefore, at least 7 of the 12 X--ray detected sources turn out to be luminous AGN.


\begin{figure*}
\psfig{file=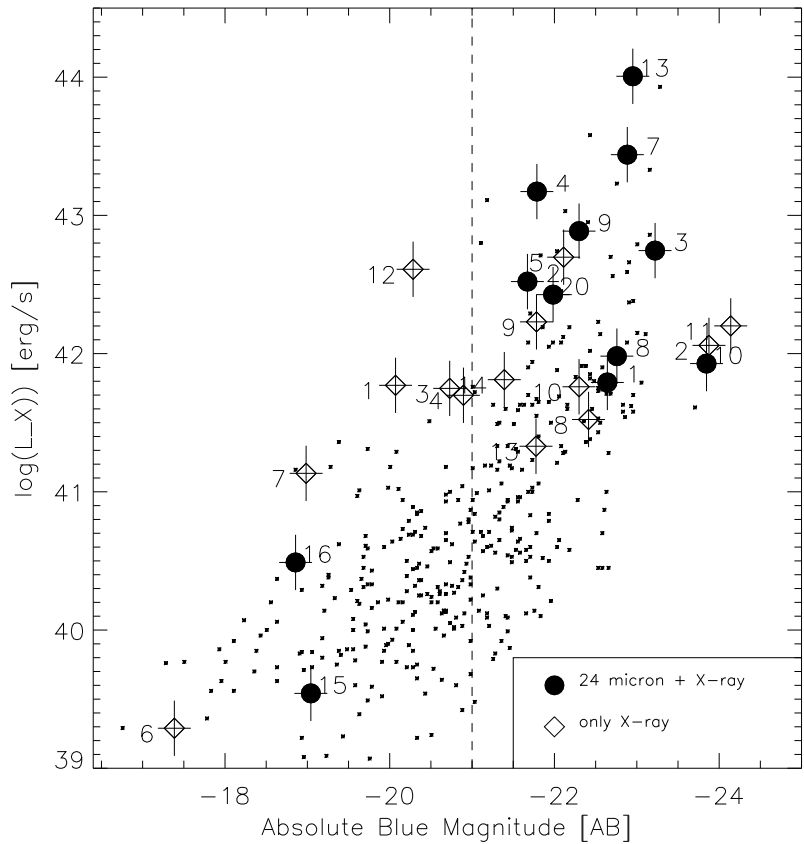,width=12cm,angle=0}
\caption{Plot of the X-ray luminosity in the 0.5-10 keV bin versus the
B-band rest-frame luminosity for the sample sources. Open diamonds are the
objects with X-ray counterparts but no detectable flux at 24 $\mu$m, filled circles
are X-ray sources with 24 $\mu$m detection. Source identification numbers refer
to the ID column in Table 3 and are reported to the left and to the right 
of the data-points for diamonds and circles, respectively.
The small symbols correspond to optical and X-ray luminosities of a local
spheroidal galaxy sample by Ellis et al. (2006).
}
\label{x_B}
\end{figure*}

\subsection{The sample of X-ray detected active spheroidal 
galaxies with no MIR emission}
\label{Xactive}

We have attempted to complement the information provided by the limited sensitivity 
of the 24 $\mu$m map by including in our analysis also the sample of 14 spheroidal 
galaxies detected in X-rays but not in the MIR.  In these cases, 
ongoing activity is traced by the excess X-ray emission.
Detailed information on these objects is reported in the third panel of 
Table 3.

The crosses in Figure \ref{fxR} indicate
the soft and hard X--ray fluxes against R magnitudes for such X-ray 
detected objects (numbers refer to the ID's reported in the third
panel of Table 3).
The sources occupy a region of the plot intermediate between that of AGNs and
that of starbursts. For 5 objects, those with ID \#3x, 5x, 9x, 11x, 12x,
an AGN predominance in the X-ray flux is suggested.

The X-ray luminosities in the 0.5-10 keV bin are reported against
the $B$-band absolute magnitude ($M_B$) as open diamonds in Figure \ref{x_B}. The comparison 
of the luminosities for the IR-detected (filled circles) and the
undetected objects (open diamonds) reveal some tendency of the former
to display larger luminosities in both bands.
Our high-$z$ spheroidal galaxies tend to show larger values of the X-ray
luminosity than the local counterparts in Figure\ref{x_B} (the local
sample of spheroidal galaxies is taken from Ellis et al., 2006, small
crosses in the Figure). In fact, the median of the X-ray luminosity
for our high-$z$ sources is on the order of $log(L_X)\sim42.2$ erg/s, while 
the median of the X-ray luminosity for the local sample is
$log(L_X)\sim40.5$ erg/s.
Even if the selection of our flux limited sample ($mag(z)<22.5$) 
should ensure the completeness at optical wavelengths, 
to account for the possibility that our sample might
be biased towards intrinsically brighter systems (simply because fainter
objects at high-$z$ could fall out of the sample), we have also
limited the comparison to the brighter systems.
If we consider sources brighter than $M_B$<-21 (as marked by the
vertical dashed line in Figure \ref{x_B}), the median of the X-ray luminosity
for our high-$z$ sources is still on the order of $log(L_X)\sim42.2$ erg/s, while 
the median of the X-ray luminosity for the local sample is brighter
than the previous case, $log(L_X)\sim41.3$ erg/s. However, also at
brighter optical luminosity the high-$z$ spheroidal population seem to
populate brighter X-ray luminosities.
However, the agreement with local data start to look better at even brighter
optical luminosities ($M_B$<-22). In this case, again $log(L_X)\sim42.2$
erg/s for the high-$z$ sample, but $log(L_X)\sim41.7$ in the local Universe.

To derive the absolute B magnitudes we have computed the observed B
magnitudes in the rest-frame, for each object, by convolving the
corresponding best-fit redshifted-SEDs with the B-band filter.
We have then converted the observed rest-frame magnitudes into their
absolute values through the corresponding luminosity distances.  
Error bars in the Figure accounts for the uncertainties on the
photometry and on the $K$-correction applied to derive the rest-frame
absolute $B$ magnitudes.

The  X-ray luminosities have been proposed to allow further rough discrimination between AGN- and
starburst-dominated phenomena: the value of $L_X \simeq 10^{42}$ erg/s
has been found to be a limit only rarely approached and never exceeded by  
star-forming galaxies (Franceschini et al. 2003; Persic et al. 2004).

An important point about the origin of the X-ray energy source is to understand
how much of it might be related with the presence of long-lived emissions by hot 
plasma coronae or by old evolved stellar populations, instead of tracing 
ongoing activity phenomena.

Discovered by the $Einstein$ Observatory (Forman et al. 1979), the hot coronae 
around local early-type galaxies were interpreted 
as plasma distributions with masses up to several $10^{10}\ M_{\odot}$
in quasi-static equilibrium in the galaxy dark matter halo. The heating and 
cooling mechanisms and timescales are not completely clear. The former, in 
particular, might be related with residual star formation and supernovae 
explosions (e.g. Forman et al. 1994).


\begin{figure*}
\psfig{file=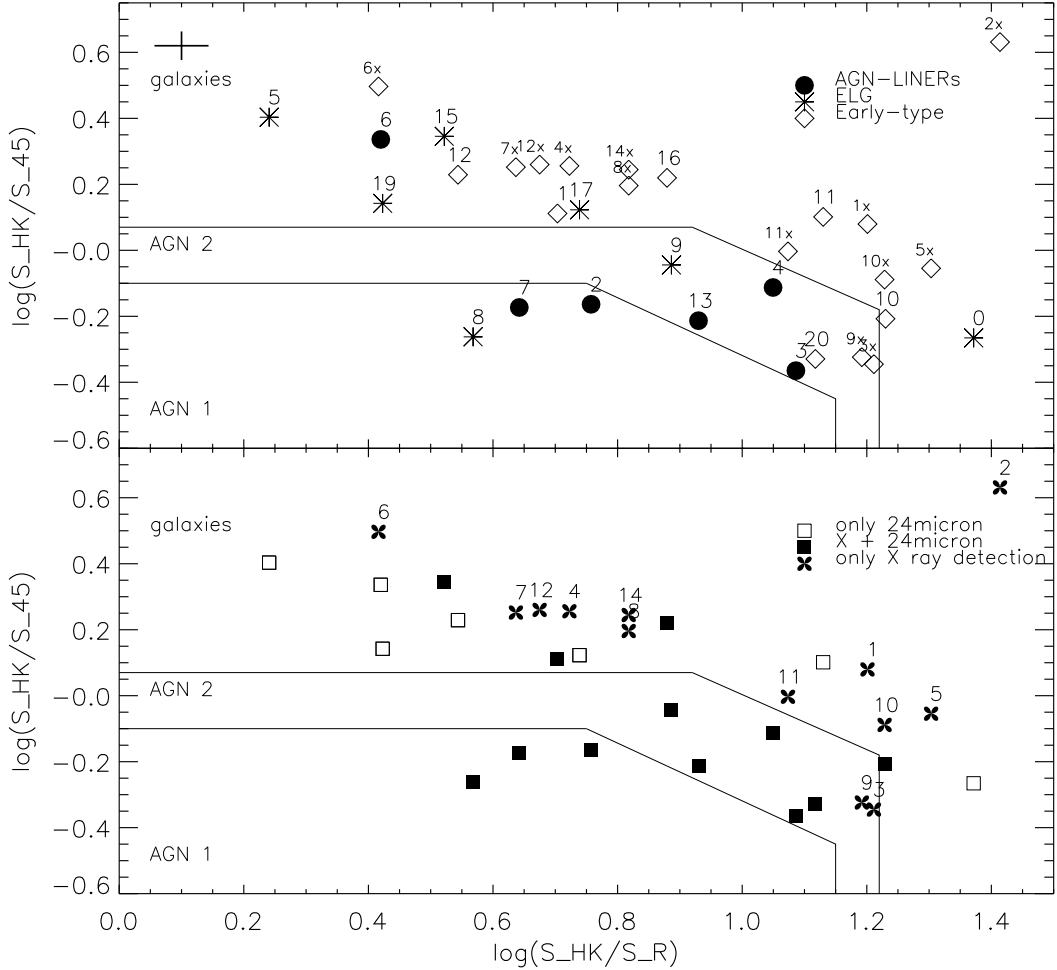,width=15cm}
\caption{
Colour-colour plot with $S_{HK}/S_{4.5}$ versus $S_{HK}/S_R$. 
$S_{HK}$ and $S_R$ are the flux densities in the near-IR $HK$-band and the optical R-band. 
The solid lines delimit regions of this plane typically populated by type-1 AGNs,  
(lower region), type-2 AGNs (middle region), while galaxies are mainly distributed in the 
upper area of the plot (Franceschini et al. 2005; note that these latter use the $H$-band
flux instead of the present $HK$, hence a small shift in the lines by an average factor
1.3 has been applied to the delimiting lines here). 
In the upper panel the different symbols indicate the spectroscopic classification 
based on the optical spectrum. In the lower panel we differentiate spheroidal galaxies 
with 24 $\mu$m flux detection only (open squares), from those detected in both MIR and 
X-rays (filled squares), and from the X-ray emitting spheroids with no detectable MIR
emission (crosses). 
For these latter the ID number is reported in the upper panel, while 
ID numbers in the lower panel refer to the sources detected in the X-rays but not in the 
MIR.
The typical error bar is indicated in the upper left angle of the top panel.
}
\label{col_franc}
\end{figure*}

\begin{figure*}
\psfig{file=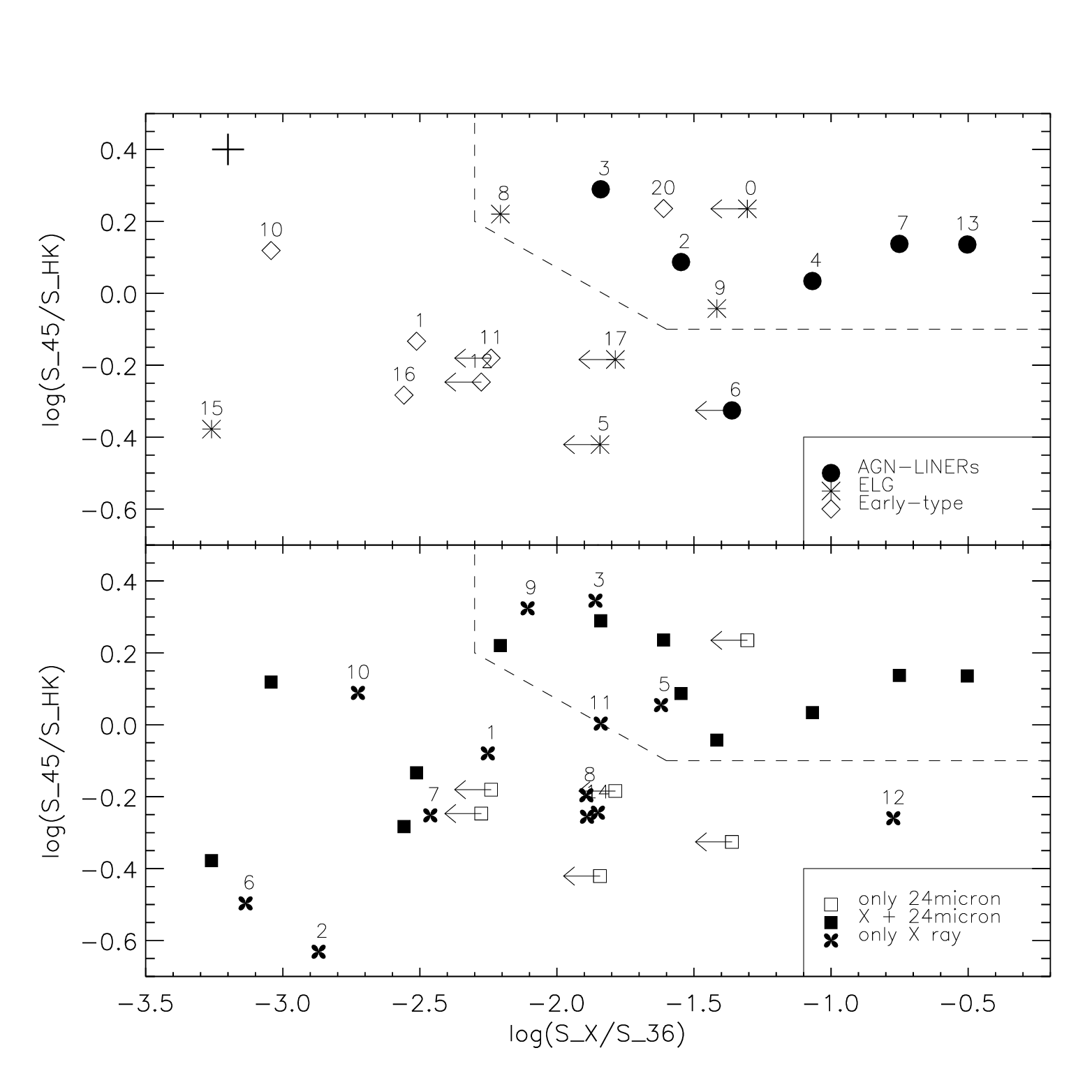,width=15cm}
\caption{
The ratio of the 4.5 $\mu$m band over the $HK'$ fluxes against the ratio of the broad-band fluxes
in the 0.5-8 keV band over the 3.6 $\mu$m band.
Different symbols correspond to different spectroscopic
classification. The meaning of the symbols is the same as reported in
Figure \ref{col_franc}.}
\label{col_franc2}
\end{figure*}


Because of these uncertainties, the interpretation of the X-ray emissions
in our high-redshift spheroidal galaxies
might turn out to be ambiguous. However, the temperatures observed in local objects 
for such gas coronae are low, $T\sim 10^7\ K$ (Forman et al. 1994), consistent with the modest 
gravitational fields confining them. For the 8 galaxies detected in hard 
(2-8 keV) X-rays, our observed X-ray fluxes require much hotter
plasmas ($T> 10^8\ K$, computed by assuming that only a single thermic
component contributes to their X-ray emission), also taking into
account the appreciable $K$-correction.
How much of the hard X-ray fluxes by early-type galaxies might be due to 
old stars in low-mass X-ray binaries (LMXB) is a problem already addressed 
in detail in a number of papers. Forman \& Jones (1989), Fabbiano et 
al. (1989) argue that only X-ray luminosities lower than $3\times 10^{40}$ erg/s
can be contributed by LMXB's, while those higher than $10^{41}$ erg/s
are dominated by gaseous emission. For our typical sample sources, the
X-ray luminosities exceeding a few times $10^{41}$ erg/s indicate that the 
LMXB contribution should be negligible, with the exception of source 
\#6x, likely contributed by either long-lived plasma emission or LMXB's.

For the other 5 objects in our sample undetected in the hard X-ray band the 
interpretation is less clear, and will be further discussed later.


\subsection{Colour-Colour Diagnostics and Broad-Band X-ray-to-IR Flux Ratios}
\label{colours}

We have looked for additional colour diagnostics to further discriminate the 
nature of the emissions in our sample of spheroids.
A combination of optical to near-IR colours has been recently suggested by 
Franceschini et al. (2005) as a diagnostic tool to disentangle among different AGN 
categories and normal galaxies. The corresponding plot for our source sample 
is reported in Figure \ref{col_franc}, showing the ratio of the $HK'$-band 
over the 4.5 $\mu$m fluxes against the $HK'$ to optical $R$-band flux ratio. 
From bottom-left to top-right, the three domains delimited by the solid 
lines correspond to the three different populations: type-1 AGNs, type-2 AGNs 
and normal/starburst galaxies, in the order.
The three regions in the figure have been empirically calibrated 
by Franceschini et al. (2005) by using a sample of sources with known
spectral types.
Type-1 AGNs and optical quasars exhibit standard blue 
optical colours (attributed to an accretion disk) and 
steeply rising near-IR spectra (due to the presence of dust re-radiation; e.g. 
Fritz et al. 2006), and are expected to populate the bottom-left corner of the 
plot. Normal or starbursting galaxies have, according to the redshift, either 
blue optical and blue near-IR colours (at low-$z$), or red optical and red
near-IR (at high-z), and tend to fall in the complementary region of the plot 
outside the solid lines.
Finally, type-2 and obscured AGNs populate intermediate colour
regions\footnote{In their original version Franceschini et al. used
standard Johnson $H$-band, instead of the $HK'$ used here: solid lines in 
Figure \ref{col_franc} have then been corrected to account for this difference.}.

In the upper panel the different symbols indicate the spectroscopic classification 
based on the optical spectrum. In the lower panel we differentiate spheroidal galaxies 
with 24 $\mu$m flux detection only, from those detected in both MIR and 
X-rays, and from the X-ray emitting spheroids with no detectable MIR
emission. 

As expected, our high-$z$ spheroidal galaxies with either IR excess or X-ray emissions 
almost completely avoid the colour region corresponding to type-1 quasars
(bottom-left corner). 

Objects with both MIR and X-ray bright detections
(filled squares in the bottom panel) dominate the type-2 AGN regions.  Top panel also reveals
that the same region is occupied by sources with characteristic AGN lines
(filled circles). Sources lacking an X-ray
counterpart (open squares) or those with an X-ray detection only (crosses) 
are mostly distributed in the region expected for normal/starforming galaxies. 
Sources detected at 24 $\mu$m and in X-rays fall mostly inside the boundaries 
of type-1 and type-2 AGNs.

We observe a generally fair agreement of this diagnostics with that of
Figure \ref{fxR}.
Source \#8, which falls in the AGN 1 region in the plot of Figure \ref{col_franc},
is only marginally consistent with an AGN classification based on its
hard-X flux and $R$-band magnitude. However, this object is spectroscopically
classified as ELG, implying a difficult interpretation of its MIR
excess. 

The near-IR colour information is complemented in Figure \ref{col_franc2}
with that on the broad-band X-ray-to-IR flux ratios, following a diagnostic scheme
also discussed in Franceschini et al. (2005). The two panels
report data with different symbols corresponding to different classification
criteria for the sources, as detailed in the inserts. The bottom panel
includes also sources with X-ray detection only (no MIR).
{\sl Bona-fide} AGNs, with clear excess fluxes in X-rays and in the MIR (4.5 $\mu$m),
occupy the region indicated in the top-right quadrant, while objects with
less extreme X-ray and IR emissivities, hence more likely dominated by 
stellar activity, are spread in the other quadrants of the plot.

We see an overall good level of agreement among these various classification
tools, with the exception of those based on the optical emission/absorption 
lines providing a classification often at variance with the others.

\section{Discussion}
\label{D}

\subsection{Local versus High-Redshift Spheroidal Galaxies}
\label{FIRex}

With few exceptions, early-morphological type galaxies in the local universe appear 
as typically inactive systems, almost completely devoid of an interstellar medium,
with no relevant star-formation or AGN activity in place.
This view has been essentially confirmed by a variety of observations, 
including those carried out with space IR and X-ray observatories.
The IRAS survey first revealed that a fraction of local
spheroidal galaxies contain some amounts of ISM gas and dust 
(Jura et al. 1987; Knapp et al. 1989). 
The {\em Infrared Space Observatory (ISO)}, with better sensitivity 
and higher resolution, has shown the presence of trace amounts of dust
and Polycyclic Aromatic Hydrocarbon molecules (PAHs) in the MIR spectra 
of a few local early-type galaxies (i.e. Madden et al. 1999; Athey et al. 
2002; Xilouris et al. 2004), results also recently confirmed by observations 
with the {\em Spitzer Space Telescope} (Pahre et al. 2004).
{\em Spitzer} IRS spectra reported by Bressan et al. (2006) show that, 
in the bulk of local early-type galaxies, the MIR emission is due to
circumstellar dust shells produced by AGB stars. 

In such local populations the dust contribution to the MIR
light is quite modest, however, providing a tiny fraction of the bolometric
galaxy emission. The corresponding amount of dust is negligible.

On the contrary, we have found that a significant fraction ($\sim$15\%) of
high-redshift spheroidal galaxies are detected with substantial mid- and 
far-IR luminosities, which are a factor $\sim$10 higher than found in the local
counterparts (see Figure \ref{fig_l12}). The large MIR excess emissions are apparent 
in the SED's of Figs. \ref{fig_sed} and \ref{fig_sed4_blend},
particularly evident if compared with the integrated spectra of local 
objects reported e.g. by Athey et al. (2002) and Bressan et al. (2006).
These emissions cannot be interpreted in terms of simple stellar mass loss. The latter
interpretation of our observed fluxes would require both mass loss rates quite larger than 
expected during any evolutionary phases of spheroidal galaxies, and metallicities of the
main stellar populations factors 2-3 higher than solar (Bressan et al. 1998, 2001).
Further evidence that circum-stellar dust by evolved stars cannot be responsible 
for the excess IR emissions is provided by the emission lines detected in 
all the optical spectra of figure \ref{specz1} 
and by the large X-ray luminosities of the majority of the sample objects.

A related question is about the origin of the observed X-ray emission for the 
sub-sample objects with no MIR detection. Less information is available for 
these spheroidal galaxies, and correspondingly more difficult is their interpretation. 
These sources tend to be less \textit{active} than the MIR emitting
ones, as for example indicated by their
typical X-ray luminosities ($log(L_X)<42$ erg/s), and by the fact that
5 out of 10 of them do not show appreciable emission lines in Figure \ref{specz}.
Here again, however, the properties for the majority of the sources 
appear to require the presence of an activity source, as 
summarized in Table 1.

\subsection{Nature of the IR and X-ray Emissions in High-z Spheroidal Galaxies}
\label{FIRex}


\subsubsection{The IR-loud Population}

With reference to the 19 spheroids with detectable MIR emission, we have found
8 objects for which all our diagnostic diagrams provide consistent evidence
for a dominant AGN presence. These are sources \#0, 2, 3, 4, 7, 9, 13 and 20.
Only four of them (\#2, 4, 7, 13) display type-2 AGN features in the optical 
spectrum. 

Source \#8 appears as a transition object in all diagrams
and the SED seems to require a dusty-torus
contribution to the MIR. Similarly, source \#6 shows a LINER
spectrum and its SED suggests an AGN contribution in the MIR.
Source \#10 also appears as a transition object, with colours 
in Figure \ref{col_franc} close to those of type-2 AGNs and a large 
X-ray luminosity in Figure \ref{x_B}.

For eight further sources (\#1, 5, 11, 12, 15, 16, 17, 19), none of our 
diagnostic diagrams indicates the presence of an AGN. For most of these, instead, 
the rest-frame $\sim$12 $\mu$m and X-ray luminosities in Figs. \ref{fig_l12} and \ref{x_B}
require the presence of star formation activity in the galaxy. Based on
our spectral best-fits to the SED data, we have computed the bolometric luminosities 
and from these estimated the star-formation rates (SFR's) using recipes discussed in 
Rowan-Robinson et al. (1997) and Elbaz et al. (2002). 
We infer values of SFR from $\sim1$ to $\sim20\ M_{\odot}/yr$. For the two low-redshift
galaxies \#5 and \#15, the SFR is instead around 0.2 $M_{\odot}/yr$ at most.


\subsubsection{The Population Emitting in X-rays only}

A basic uncertainty is whether an activity (AGN or starburst)  is present at all in these
galaxies, or if their X-ray flux might be entirely explained by long-lived
emissions by hot plasma coronae or LMXB populations.
Eight out of the 14 X-ray--only galaxies are detected in the hard X-ray 
(2-8 keV) band with luminosities larger than $10^{41}$ erg/s.
As discussed in Sect. \ref{Xactive}, the combination of the large X-ray 
luminosity and hard X-ray band implies either an AGN or an ongoing starburst
to reside in these sources. Of these, at least 5 (ID \#3x, 5x, 9x, 11x, 12x)
show evidence for an  AGN in the large value of the X-ray luminosity 
(Figure \ref{col_franc2}), the large X/O ratios (Figure \ref{fxR}), 
and in the optical spectra (Figure \ref{specz}).
Sources \#4x and \#14x are likely to host a luminous 
starburst,
while source \#7x is a low-optical luminosity, moderately active galaxy in X-rays,
and could host a low-luminosity AGN.

Two of the remaining 6 objects detected only in the soft X-rays
(ID \#1x and \#13x), show prominent [OII] and other emission lines and are
star forming galaxies.
Source \#2x is a very luminous soft X-ray and very bright galaxy in the optical
(Figure \ref{x_B}), with a passive (absorption-line, post-starburst) optical spectrum,
and likely to host an obscured starburst, or post-starburst young stellar
population.

Source \#6x is, on the contrary, a very low-luminosity local object in both 
soft X-rays and the optical. The X-ray emission is likely to originate
from a hot plasma corona, also confirmed by the complete absence of
optical emission lines.

Finally, for the soft X-ray emitting objects \#8x and \#10x we lack an optical
spectrum. Both are moderately luminous in the optical and X-rays
($M_{B}\simeq-22$, $L_X\simeq 5\times 10^{41}$ erg/s) and all their 
properties are consistent with those of starbursts.

\subsection{Galaxy Morphologies}
\label{galmorph}

The unique multi-band ACS imaging add valuable physical information on the sample
galaxies. Indeed, if the \textit{activity} is triggered by merging or interactions,
we expect we would be able to detect it.

The images in Figs. \ref{s_morph} and \ref{blend} first confirm that the far-IR emitting
objects appear to have been detected in a higher \textit{activity} stage than the
X-ray emitting only.  Of the 19 IR-loud galaxies, 12 show morphological peculiarities
or some evidence of interactions, while only 4 of the 14 sources detected in X-rays
show morphological irregularities.

In some cases (ID \#4, \#8, \#9, \#10, \#11, \#9x) the spheroidal galaxies look part of 
a group.

However, the majority of the X-ray emitting and a good fraction of the IR-loud
objects (i.e. 15 out of 33 in the total sample) appear well isolated and showing
the regular elliptical contours of spheroidal galaxies.

\subsection{Statistical Considerations}
\label{stat}

From the combined diagnostics considered in this work
(morphology, spectroscopy, X-ray emissions, colours, broad-band luminosity 
ratios), a rather heterogeneous scenario emerges concerning the nature of the
population of "active" spheroidal galaxies at redshifts from 0.3 to 1.2.

In summary, among the 33 spheroids with excess MIR or X-ray emissions,
13 show clear evidences for the presence of an obscured AGN as mainly 
responsible. For 16 objects (8 among the IR-loud and 8 more in the X-ray
loud classes) the most likely interpretation is in terms of an
hidden starburst, with typical SFR values between 1 and few tens $M_{\odot}/yr$.
The remaining 4 galaxies display less characteristic, intermediate 
physical properties.  Among them, the rate of star-formation appears to be modest 
in the 2 lowest-redshift galaxies (\#5 and 15) or even absent in \#6x.

In conclusion, almost 20\% of the optically-selected population of faint
high-$z$ spheroidal galaxies show evidence of activity at the instant
of the observation.  Roughly half of them ($\sim$10\% of the total optically-selected
spheroidal galaxy population), appear dominated by AGN, the other half
by starburst activity (another $\sim$10\% of the total). In either case, the
duration of the phenomenon is expected to be short, of the order of 
$10^8\ yr$ (e.g. Forster-Schreiber et al. 2003; Cavaliere \& Padovani 
1989), and likely to follow merging or interaction with a gas-rich 
system, able to funnel fresh gas into the galaxy nucleus. The process 
would then be expected to contribute new stars and to increase the 
nuclear super-massive black hole in the AGN sub-class.

Compared with the 5--7 Gyrs corresponding to the Hubble times at the 
source redshifts, the expected short duration of the processes and the 
significant fractions of the objects detected in the active phase 
are consistent with the fact that the whole spheroidal population
undergoes few to several such episodes at $z\leq1$.

Then assuming for example as a reference for the SFR the value of 10 $M_{\odot}/yr$ 
and a star-formation timescale of $10^8\ yr$, this would imply several 
such episodes for each galaxy, each one contributing of the order of few percent 
of the final stellar content of the galaxy on average.  For the most
massive galaxies this is not expected to modify significantly the stellar
mass below $z\sim1$, while for objects with $M<10^{11}\ M_{\odot}$ 
the increase with cosmic time would be expected to be more substantial.

Altogether, these results are at least not inconsistent with recent
analyses of the evolution of the galaxy stellar mass as a function of redshift
based on complete near-IR selected samples with spectroscopic follow-up
(Dickinson et al. 2003; Bundy et al. 2005; Fontana et al. 2004; 
Franceschini et al. 2006).  
Moreover, starting from a similar (and partially overlapping) morphological sample as used in this paper, 
Treu et al. (2005) confirmed the presence of star formation in spheroids at high 
redshift. They studied the spectroscopic properties of spheroids and the evolution of 
the fundamental plane, and confirm a scenario in which the efficiency of SF is 
enhanced in the most massive galaxies at the higher redshifts, and proceeds with 
longer timescales in the smaller ones (the $downsizing$ effect).

Our results, in which AGN and starburst activities are tightly intertwined
and actually almost inextricable in several of our galaxies, emphasize
the continuity of the two processes and further support the case for a 
co-evolution of quasars and bulges (Franceschini et al. 1999; Granato et al. 2004). 


\section{Conclusions}
\label{concl}

To complement searches for recent or ongoing star-formation and nuclear 
accretion in high-redshift spheroidal galaxies, we have exploited deep 
MIR and X-ray observations by $Spitzer$ and $Chandra$ in the GOODS-N area.
This allowed us to identify signs of hidden activity that might escape optical
investigations based on colours or strong morphological signatures of merging or 
interactions.

Our reference sample are 168 morphologically classified spheroidal 
(elliptical/lenticular) galaxies with $z_{AB}<22.5$ selected from GOODS ACS imaging.
We have found that a significant fraction of the E/S0's show
unexpected emissions in the MIR:
nineteen of them have 24-$\mu$m detections in the GOODS catalogue, 
while 6 have detectable radio flux at 1.4 GHz with $>$40 $\mu$Jy. 

We have also looked for traces of hidden activity in our optically-selected 
spheroidal population in the 2 Ms $Chandra$ catalogue and detected 
significant flux from 25 sources (13 of which having 2-8 keV hard X-ray 
emission). Of these, 11 belong to the MIPS 24-$\mu$m catalogue and 14 more
are detected in X-rays only.

The nature of the near- and MIR emission observed by $Spitzer$ 
is analyzed through the SEDs based 
on the available multi-wavelength photometry, including X-ray, UV, optical,
near-, MIR, and radio fluxes. 
The inferred amount of diffuse dust appears substantially in excess 
of that expected by mass loss from late-type stars. 
We have used in addition a variety of diagnostics, including optical spectroscopy,
colour-colour diagrams, and broad-band flux ratios to interpret the nature of
the energy source.

When the available independent diagnostics are compared, in general they provide
consistent classifications about the nature of the activity in the
spheroidal population.
However, the SED fitting study alone tends to privilege an AGN
predominance in almost all sources. X-ray diagnostics (luminosities
and spectral analysis) and colour-colour diagrams instead provide much
correlated classifications. Given that, in principle, none of these
diagnostics alone can be considered as conclusive, only trough a panchromatic
comparison of them we can reach an accurate comprehension of the
underlying physical processes.     

Our multi-wavelength analysis of the X-ray and MIR properties leads us to 
conclude that at least 8 of the 19 IR-detected sources should hide an obscured AGN, while
the X-ray undetected sources are more likely dominated by ongoing star formation.

More uncertain is the interpretation of spheroidal galaxies detected in the X-ray bands
only and undetected in the MIR, a basic question being to understand whether 
an activity source is present at all, 
or if the X-ray fluxes might be entirely explained by long-lived
emissions by hot plasma coronae or LMXB populations.
Our analysis shows evidence for the presence of an AGN or an ongoing starburst
in the majority of them.

We conclude that 15\% to 20\% of the original sample of 168 spheroidal galaxies in 
GOODS-N are detected during phases of prominent activity, probably following an event of 
merging/interaction. Our morphological analysis has revealed that only roughly a 
half of the galaxies show evidence of some peculiarities, the other half displaying
standard spheroidal morphologies. This might be taken as evidence that our
IR/X-ray selection tends to pick up the latest stages of a merging/interaction
process.
 
Thirteen galaxies (8\% of the complete sample) show evidence for the presence of an 
obscured AGN, and for other 16 (10\%) the most likely interpretation is in terms 
of a hidden starburst.
Due to the expected short lifetimes of these IR and X-ray emissions, 
this observed fraction implies widespread activity in this class of galaxies 
during the cosmic epochs -- $z\sim 0.3$ to $z\sim 1$ -- subject to our investigation.

\section*{Acknowledgments}

This work is based on observations made with the {\it $Spitzer$ Space Telescope},
which is operated by the Jet Propulsion Laboratory, California Institute of
Technology under NASA contract 1407.
Support for this work, part of the $Spitzer$ Space Telescope Legacy Science
Program, was provided by NASA through an award issued by the Jet Propulsion
Laboratory, California Institute of Technology under NASA contract 1407.

ACS was developed under NASA contract NAS 5-32865, and this research
has been supported by NASA grant NAG5-7697. We are grateful for an
equipment grant from  Sun Microsystems, Inc.
The Space Telescope Science
Institute is operated by AURA Inc., under NASA contract NAS5-26555.

We thank an anonymous referee for a careful reading of the paper and
numerous useful comments.
We also thank Marcella Brusa for useful discussions and Mary Polletta
for providing us with her SED templates..

We acknowledge financial contribution from contract ASI-INAF I/023/05/0.

\bsp 


\begin{landscape}
\begin{table}
\begin{minipage}{200mm}

\caption{Radio, MIR, NIR, Optical and X-ray Properties}
\tiny
  \begin{tabular}{rrrrrrrrrrrrrrrrrrrrrr} 

& & 
\\\hline \hline
\\
\multicolumn{1}{c}{ID} & \multicolumn{1}{c}{AID} &\multicolumn{1}{c}{ RA} &\multicolumn{1}{c}{ DEC} & \multicolumn{1}{c}{z} & \multicolumn{1}{c}{f$_{1.4GHz}$} & \multicolumn{1}{c}{f$_{24}$} & \multicolumn{1}{c}{f$_{16}$}& \multicolumn{1}{c}{f$_{8.0}$} &\multicolumn{1}{c}{ f$_{5.8}$} &\multicolumn{1}{c}{f$_{4.5}$} &\multicolumn{1}{c}{f$_{3.6}$} & \multicolumn{1}{c}{f$_{HK^{'}}$} & \multicolumn{1}{c}{f$_z$} & \multicolumn{1}{c}{f$_i$} &\multicolumn{1}{c}{f$_R$}  & \multicolumn{1}{c}{f$_V$} & \multicolumn{1}{c}{f$_B$} &\multicolumn{1}{c}{ f$_U$} & \multicolumn{1}{c}{f$_{0.5-2 keV}$} & \multicolumn{1}{c}{f$_{2-8 keV}$}\\ 
& &\multicolumn{1}{c}{ (hours)} &\multicolumn{1}{c}{ (deg)} & &\multicolumn{1}{c}{ $\mu$Jy} &\multicolumn{1}{c}{ $\mu$Jy} &\multicolumn{1}{c}{ $\mu$Jy} &\multicolumn{1}{c}{ $\mu$Jy} &\multicolumn{1}{c}{ $\mu$Jy} & \multicolumn{1}{c}{$\mu$Jy} &\multicolumn{1}{c}{$\mu$Jy} &\multicolumn{1}{c}{$\mu$Jy} &\multicolumn{1}{c}{$\mu$Jy} &\multicolumn{1}{c}{$\mu$Jy} &\multicolumn{1}{c}{$\mu$Jy} &\multicolumn{1}{c}{$\mu$Jy} & \multicolumn{1}{c}{$\mu$Jy} & \multicolumn{1}{c}{$\mu$Jy}  & \multicolumn{2}{c}{erg s$^{-1}$ cm$^{-2}$} \\
\\\hline 
\multicolumn{21}{c}{24-$\mu$m non-blended sources}\\ \hline
\\
  1 &67 & 189.01359558 &   62.18645859   &0.638  &124 &1210.0 &655  &   92.41 &  97.56  &  83.64 &  159.99 &   94.60 &  38.09   &   31.77  &   21.41  &   8.46  &   5.02  &  3.83   &    1.42e-16 &  $<$3.21e-16  \\	
  2 &82 & 189.03387451 &   62.17671967   &0.679  &217 &2300.0 &$-$  &  266.70 & 124.82  &  71.61 &   79.97 &   48.77 &  19.64   &   18.09  &    8.59  &   3.22  &   1.41  &  1.03   &    1.37e-16 &   1.80e-15  \\    
  3 &110 & 189.07011414 &   62.10406113   &1.141  & $-$ & 82.0 &$-$  &  34.86  & 43.53   & 58.50  &  73.26  &   $-$   & 10.35    &   4.84   &   2.07   &  0.70   &  0.43   & 0.26    &   3.38e-16  & $<$1.14e-15  \\	
  5 &$-$  & 189.07263184 &   62.23794937   &0.156  & $-$ & 89.8 &$-$  &  57.61  & 15.06   & 16.92  &  28.41  &  37.08  & 27.39    &  27.79   &  24.62   & 19.93   & 12.08   & 8.62    & $-$           &  $-$       \\    
  6 &$-$  & 189.07391357 &   62.13871002   &0.512  & $-$ &123.0 &$-$  &   9.57  &  4.29   &  5.32  &   4.92  &   9.36  &  5.07    &   5.12   &   4.38   &  3.13   &  2.30   & 2.54    &   $-$       &  $-$       \\    
  7 &115 & 189.07490540 &   62.27656174   &0.680  & 47  &664.0 &$-$  &  95.12  & 87.88   & 94.97  & 105.45  &  57.58  & 27.26    &  24.20   &  14.53   &  8.81   &  5.36   & 4.31    &   4.67e-15  &  1.05e-14  \\    
  8 &149 & 189.11378479 &   62.21617126   &1.223  & $-$ &138.0 &$-$  &  15.60  & 14.65   & 21.14  &  24.69  &  10.59  &  6.08    &   4.36   &   3.13   &  2.10   &1.12$^{*}$& $-$    &   5.05e-17  & $<$1.36e-16 \\	
  9 &160 & 189.12226868 &   62.27042007   &0.848  & $-$ &208.0 &125  &  54.13  & 57.20   & 60.72  &  90.56  &  48.37  & 19.58    &  14.76   &   6.18   &  2.76   &  1.36   & 1.33      &  1.15e-16  &  2.59e-15 \\     
 10 &169 & 189.13232422 &   62.15205002   &0.845  & $-$ &277.0 &$-$  & 183.81  &178.80   &260.76  & 389.44  &   $-$   &28.84$^{*}$&14.52$^{*}$&  $-$   &3.44$^{*}$& $-$    & $-$       &  9.76e-17  & $<$2.66e-16  \\	 
 12 &$-$  & 189.19013977 &   62.32765961   &0.277  & $-$ & 91.1 &$-$  &  24.14  & 26.11   & 35.21  &  40.21  &  51.71  & 27.26    &  22.96   &  17.06   &  9.33   &  2.97   & 1.17    &   $-$       &  $-$      \\     
 13 &240 & 189.19314575 &   62.23471069   &0.961  &179  &211.0 &130  &  92.92  & 79.74   & 77.76  &  93.96  &  47.33  & 16.91    &  11.05   &   5.59   &  3.35   &  1.50   & 1.07    &   3.26e-15  &  2.17e-14 \\     
 15 & 383  & 189.31648254 &   62.19979858   &0.105  & $-$ &221.0 &$-$  &  99.58  &126.59   &190.62  & 310.11  & 378.43  &218.31    & 176.33   & 127.03   & 80.42   & 30.29   & 12.38   &  5.33e-17  & $<$1.94e-16 \\      
 16 &388 & 189.31832886 &   62.25347900   &0.230  &187  &219.0 &$-$  &  72.49  & 68.01   & 95.11  & 133.84  & 111.30  & 35.57    &  26.61   &  15.26   &  6.22   &  2.59   & 1.66      &  8.96e-17  & $<$1.83e-16  \\     
 17 & $-$ & 189.31959534 &   62.21960068   &0.899  & $-$ & 86.5 &$-$  &   3.61  &  7.61   &  9.28  &  13.57  &  11.80  &  5.17    &   4.25   &   2.25   &  1.61   &  1.13   & 1.27    &    $-$       &  $-$      \\     
 19 &$-$  & 189.42692566 &   62.30044937   &0.278  & $-$ & 97.5 &$-$  &  22.08  &  9.34   & 11.40  &  11.65  &  14.43  &  8.90    &   7.56   &   5.98   &  3.98   &  1.82   & 1.27    &    $-$	 &  $-$       \\    
 20 &468  & 189.45289612 &   62.22037125   &0.911  & $-$ &398.0 &$-$  & 130.19  & 78.22  &  47.66  &  36.20  &  23.04  &  7.38    &   4.73   &   1.70   &  0.68   &  0.19   & 0.19    &   $<$6.91e-17  	 & 6.7e-16    	  \\
\\ \hline
\multicolumn{21}{c}{24-$\mu$m blended sources} \\ \hline
\\
  0 &$-$  &  188.96078491&   62.16981125   & 0.410 &  $-$& 190.0&$-$  &  13.10  & 15.05   & 17.34  &  15.48  &   8.40  &0.95$^{*}$&0.52$^{*}$& $-$      &0.34$^{*}$&  $-$   & $-$     &     $-$      &   $-$ \\      
  4 & 113& 189.07121277 &   62.16996002   &0.845  & 62  & 88.2 &$-$  &  19.79  & 34.77   & 44.58  &  69.45  &  34.28  &  9.65    &   6.09   &   3.07   &  1.30   &  0.94   & $-$       &  1.59e-15  &  3.11e-15  \\    
 11 &$-$  & 189.15679932 &   62.14453888   &0.766  & $-$ & 82.3 &$-$  &  10.95  & 20.03   & 22.09  &  37.38  &  27.83  &  7.24    &   4.99   &   2.07   &  0.80   &  0.27   & 0.13    &   $-$       &  $-$       \\     
\\ \hline
\multicolumn{21}{c}{X-ray detected sources lacking a 24-$\mu$m counterpart} \\ \hline
\\

 1x  & 414  &  189.35189819   &   62.21170044   &  0.798 & $-$  & $-$  & $-$  &  13.41   & 22.69    &   30.13   &   48.44    & 33.121   &   7.974   &   5.544   &  2.278     &  0.902     & 0.251    &  0.197    &    9.49e-17   &   <2.53e-16     \\  
 2x  & 230  &  189.18499756   &   62.19263077   &  1.013 & $1290$  & $-$  & $-$  &  48.31   & 91.71 &   21.08   &  216.79    & 82.543   &  19.162   &   9.354   &  3.479     &  0.849     & 0.228    &  0.100    &    7.74e-17   &   <1.80e-16   \\    
 3x  & 294  &  189.23709106   &   62.21722031   &  0.474 & $49$  & $-$  & $-$  &  18.80   & 29.62   &   49.20   &   66.07    & 20.337   &   4.402   &   2.333   &  1.366     &  0.776     & 0.593    &  0.573    &    3.14e-16   &    4.31e-16     \\  
 4x  & 249  &  189.20028687   &   62.21926880   &  0.475 & $-$  & $-$  & $-$  &  19.40   & 36.57    &   42.72   &   62.89    & 70.450   &  28.790   &  22.499   & 14.583     &  5.156     & 1.653    &  0.514    &    2.37e-16   &    4.48e-16     \\  
 5x  & 286  &  189.23100281   &   62.21989059   &  0.954 & $-$  & $-$  & $-$  &  18.69   & 30.77    &   42.38   &   61.55    & 34.193   &   8.896   &   4.843   &  1.859     &  0.673     & 0.198    &  0.067    &    4.08e-16   &    7.75e-16      \\ 
 6x  & 257  &  189.20588684   &   62.22978973   &  0.089 & $-$  & $-$  & $-$  &   46.18  &  73.60   &   115.32  &   185.35   & 331.150  &  214.763  &  181.691  & 138.798    &  86.179    & 35.696   &  14.557   &     4.22e-17  &    <1.58e-16      \\ 
 7x  & 274  &  189.22029114   &   62.24567032   &  0.321 & $168$  & $-$  & $-$  &  47.50   & 70.37    &  124.61   &  160.96    &203.698   &  87.793   &  75.637   & 51.387     & 22.824     & 6.772    &  2.140    &    2.44e-16   &    2.19e-16     \\  
 8x  & 354  &  189.28599548   &   62.25051117   &  0.569 & $-$  & $-$  & $-$  &   7.64   & 16.88    &   19.10   &   27.36    & 27.447   &  10.111   &   7.814   &  4.563     &  1.595     & 0.484    &  0.197    &    8.50e-17   &   <2.09e-16      \\ 
 9x  & 241  &  189.19331360   &   62.25815964   &  0.850 & $-$  & $-$  & $-$  &  40.55   & 53.45    &    55.46   &  84.72    & 24.077   &   6.803   &   4.533   &  1.691     &  0.630     & 0.145    &  0.132    &    1.91e-16   &    3.38e-16      \\ 
10x  & 210  &  189.16569519   &   62.26335907   &  0.848 & $-$  & $-$  & $-$  &   32.37  &  55.22   &    68.80  &   121.24   &  51.343  &   15.605  &   10.047  &   3.315    &   1.084    &  0.231   &   0.057   &     9.12e-17  &    <1.68e-16      \\ 
12x  & 212  &  189.16729736   &   62.28221893   &  0.943 & $-$  & $-$  & $-$  &  11.98   & 16.59    &   26.21   &   33.46    & 23.806   &   9.674   &   6.587   &  2.197     &  0.914     & 0.329    &  0.249    &    1.30e-16   &    2.73e-16      \\ 
12x  & 194  &  189.15321350   &   62.19900894   &  0.556 & $-$  & $-$  & $-$  &   10.47   &  9.03    &    13.30  &   26.81    & 22.150   &   9.472   &   7.481   &  5.118     &  2.725     & 1.781    &  0.940    &    1.62e-15   &    2.01e-15      \\ 
13x  & 131  &  189.09410095   &   62.21123123   &  0.632 & $-$  & $-$  & $-$  &   0.0   & 34.04    &    21.68   &   55.46    &  0.000   &   0.000   &   0.000   &  0.000     &  0.000     & 0.000    &  0.000    &    2.47e-17   &   <1.96e-16       \\  
14x  & 114  &  189.07359314   &   62.22906113   &  0.534 & $-$  & $-$  & $-$  &   15.22   & 27.222    &    35.97   &   56.22    & 57.806   &  22.204   &  17.204   &  9.611     &  3.043     & 0.952    &  0.260    &    3.77e-17   &    6.59e-16     \\ 
\\ \hline \hline

\end{tabular}
\footnotesize
$^{*}$ ACS data from Bundy et al. (2005)

\end{minipage}

\label{tab_cat}
\end{table}
\end{landscape}

\end{document}